\documentclass[nofootinbib,%
 reprint,
 superscriptaddress,
preprintnumbers,
 amsmath,amssymb,
 aps,
prx
]{revtex4-2}

\usepackage{float}
\usepackage{bbold}
\usepackage{braket}
\usepackage{amssymb,amsmath}
\usepackage{tikz}
\usepackage{bbm}
\usepackage{slashed}
\usepackage{svg}
\usepackage{microtype}
\usepackage{amsfonts}
\usepackage{array}
\usepackage{enumerate}

\usepackage{graphicx}
\usepackage{bm}
\usepackage[colorlinks, citecolor=blue,anchorcolor=red,menucolor=red,linkcolor=red,filecolor=red,runcolor=red,urlcolor=blue,frenchlinks=red]{hyperref}
\hypersetup{breaklinks=true}
\usepackage[nameinlink]{cleveref}
\crefformat{equation}{Eq.~(#2#1#3)}
\crefformat{figure}{Fig.~#2#1#3}
\crefformat{section}{Sec.~#2#1#3}

\usepackage[normalem]{ulem}

\graphicspath{{./figures/}{./}}

\begin{document}
\preprint{RIKEN-iTHEMS-Report-26}

\title{Detecting Multiple Phase Transitions in Lattice Systems with Intrinsic Dimensions}

\author{Jie Mei}
\email{meijie@ucas.ac.cn}
\affiliation{School of Nuclear Science and Technology, University of Chinese Academy of Sciences, Beijing, 100049,  China}

\author{Tetsuo Hatsuda}
\email{thatsuda@riken.jp}
\affiliation{RIKEN Interdisciplinary Theoretical and Mathematical Sciences (iTHEMS), Wako, Saitama 351-0198, Japan}
\affiliation{Kavli Institute for the Physics and Mathematics of the Universe (Kavli IPMU), WPI,  
The University of Tokyo, Kashiwa, 
277-8568, Japan}%

\author{Mei Huang}
\email{huangmei@ucas.ac.cn}
\affiliation{School of Nuclear Science and Technology, University of Chinese Academy of Sciences, Beijing, 100049,  China}

\author{Lingxiao Wang}
\email{lingxiao.wang@riken.jp}
\affiliation{RIKEN Interdisciplinary Theoretical and Mathematical Sciences (iTHEMS), Wako, Saitama 351-0198, Japan}
 \affiliation{Institute for Physics of Intelligence, The University of Tokyo, Hongo,  Tokyo 113-0033, Japan}


\begin{abstract}
Lattice systems with multiple nearby transitions pose two related challenges: resolving distinct transition scales and identifying the degrees of freedom primarily associated with each transition. We show that the intrinsic dimension of Monte Carlo configuration ensembles, estimated by the two-nearest-neighbors method, provides a geometric diagnostic for both problems. In the two-dimensional $q$-state clock model, the intrinsic dimension distinguishes the ordered, quasi-critical, and disordered regimes for both well-separated ($q=9$) and closely spaced ($q=5$) Berezinskii--Kosterlitz--Thouless transitions. In the $q=5$ case, the
intermediate phase appears as a broad low-dimensional valley even when energy and magnetization do not separately resolve the two transitions. In the four-dimensional $U(1)$ Higgs model, we introduce channel-decomposed intrinsic dimensions based on gauge-invariant plaquette and Higgs variables. The dominant response of each channel tracks transitions associated with the corresponding degrees of freedom, while the combined channel retains features of both. We further show that intrinsic dimensions evaluated directly on gauge-variant fields are dominated by gauge-orbit directions, demonstrating the importance of removing gauge redundancy before interpreting configuration-space geometry. These results establish channel-decomposed intrinsic dimension as a geometric probe of lattice systems with multiple transitions and motivate its application to disentangling deconfinement and chiral crossover scales in full QCD. \end{abstract}
\maketitle

\section{Introduction}

Finite-temperature QCD is characterized by two closely related phenomena, chiral symmetry restoration and deconfinement~\cite{Polyakov:1977bh,McLerran:1981pb,Svetitsky:1982gs,Karsch:2001cy,Fukushima:2010bq}. In the chiral limit, the chiral condensate serves as an order parameter for spontaneous chiral-symmetry breaking~\cite{Banks:1979yr,Pisarski:1983ms}. In pure gauge theory, the Polyakov loop is an exact order parameter associated with center symmetry~\cite{Polyakov:1977bh,McLerran:1981pb,Svetitsky:1982gs}. With dynamical quarks of finite mass, however, both the chiral and center symmetries are explicitly broken, and lattice QCD at physical quark masses exhibits a smooth crossover rather than a genuine thermodynamic phase transition~\cite{Aoki:2006we,Borsanyi:2010bp}. One may nevertheless define pseudo-critical temperature scales from chiral and gluonic observables. Whether these observables identify a common crossover scale or nearby but distinct scales remains an open question~\cite{Bazavov:2011nk,HotQCD:2018pds,Borsanyi:2020fev}.  For a comprehensive discussion of the QCD phase diagram and related lattice results, see Refs.~\cite{Fukushima:2010bq,Ratti:2018ksb,Aarts:2023vsf}.

This motivates the search for alternative geometric diagnostics that do not require a preselected order parameter. High-dimensional datasets typically lie on or near low-dimensional manifolds whose effective dimensionality, the intrinsic dimension ($I_d$), encodes nontrivial geometric information about the data. Methods such as principal component analysis (PCA)~\cite{Pearson:1901pca}, variational autoencoders (VAE)~\cite{Kingma:2013hel}, maximum likelihood estimation~\cite{Levina:2004mle}, fractal-based methods~\cite{Camastra:2002est}, and geodesic-entropic-graph techniques~\cite{Costa:2004geo,Granata:2016acc} extract this dimensionality without strong parametric assumptions. In parallel, topological data analysis (TDA) provides a framework for extracting topological features and characterizing phase structure directly from lattice QCD configurations~\cite{Edelsbrunner:2000tps,Carlsson:2009zke,Sale:2022qfn,Spitz:2022tul}. In lattice field theory, the configuration space sampled by Monte Carlo simulations likewise has an effective dimensionality far below the naive embedding dimension, and this dimensionality evolves as external parameters such as temperature are varied. 
At a continuous phase transition, the emergence of long-range correlations reorganize the probability distribution of field configurations controlled by a few critical exponents~\cite{Zinn-Justin:2002qft,Cardy:1996sr}.
Previous studies have shown that this reorganization can generate characteristic features in intrinsic-dimension estimators, including minima, plateaus, or changes of slope depending on the system and on the channel considered~\cite{Mendes-Santos:2020msa}. The $I_d$ can therefore serve as an unsupervised quantity for detecting phase transitions even without a preselected physical observable. This perspective has been developed in recent works employing the two-nearest-neighbors (2NN) estimator~\cite{Facco:2017est} to identify critical behavior in classical and quantum many-body systems~\cite{Mendes-Santos:2020msa,Mendes-Santos:2021mhs,Panda:2023zcj}, and related geometric diagnostics based on configuration-space distances have likewise been used to probe phase transitions~\cite{Su:2024oqd}. 

These geometric diagnostics sit within a broader use of machine learning in studies of phase transitions. Alongside supervised~\cite{Carrasquilla:2016oun,Broecker:2017hjl,vanNieuwenburg:2016zsd} and unsupervised~\cite{Wang:2016nmn,Wetzel:2017olt,Hu:2017wey} approaches, there is growing interest in applications to lattice field theories~\cite{Boyda:2022nmh,Zhou:2023pti,Tomiya:2025quf, Aarts:2025gyp}. Beyond lattice configurations, the $I_d$ of data representations has also been studied in deep neural networks~\cite{Ansuini:2019idn} and large language model inferences~\cite{Ruan:2026mdn}, and efficient software implementations are readily available~\cite{Glielmo:2022dad}.
 
For the $I_d$ to be a useful diagnostic in QCD-like systems, it must demonstrate two distinct capabilities. The first is to resolve closely-spaced transitions, as may occur when deconfinement and chiral restoration take place at nearby but not identical temperatures. The second is to disentangle transitions associated with different sectors of degrees of freedom. 
Since deconfinement is primarily reflected in gluonic observables whereas chiral restoration is diagnosed through fermionic observables, the two phenomena probe different physical sectors of the theory. Conventional bulk observables can fail on either count. In this work, we test these two capabilities separately with two benchmark systems chosen to isolate each challenge.
 
The first capability, resolving closely-spaced transitions in a single-field system, is tested with the two-dimensional $q$-state clock model. The clock model is a discretization of the XY model in which continuous spin orientations are restricted to $q$ equally-spaced directions. For $q\leq 4$,
the model exhibits a single continuous transition, while for $q>4$ it exhibits two Berezinskii--Kosterlitz--Thouless (BKT) transitions~\cite{Berezinskii:1970pzv,Kosterlitz:1973xp,Kosterlitz:1974nba} at temperatures $T_1<T_2$, with an intermediate quasi-long-range-ordered phase ~\cite{Jose:1977gm,Elitzur:1979uv}. The separation $T_2-T_1$ shrinks monotonically as $q$ decreases toward 4, and grows as $q$ increases toward the XY limit $q\rightarrow\infty$~\cite{Tobochnik:1982qsc,Lapilli:2006uni,Hwang:2009six,Kumano:2013rsp,Gupta:1992zz,Surungan:2019hjf}. By tuning $q$ one can therefore control the proximity of the two transitions in a well-characterized setting, and ask how the $I_d$ signal degrades as they approach each other and what complementary information it carries relative to conventional bulk observables. We contrast a well-separated case ($q=9$) with a closely-spaced one ($q=5$).
 
The second capability, identifying which degree of freedom drives a given transition, is tested with the four-dimensional $U(1)$ Higgs model. This model contains both a compact gauge field~\cite{Wilson:1974sk,Polyakov:1976fu} and a charged scalar field, with a Fradkin--Shenker phase diagram exhibiting confinement, Coulomb, and Higgs regions separated by gauge, Higgs, and $Z_2$ transition lines~\cite{Fradkin:1978dv,Osterwalder:1977pc}. Early numerical studies mapped out the phase structure in detail~\cite{Callaway:1981rt,Jansen:1985nh}; for a modern review of confinement and the Higgs--confinement distinction, see Ref.~\cite{Greensite:2011zz}. Related phase transitions in discrete $Z_N$ gauge models have also been studied~\cite{Borisenko:2013sqa}. We introduce a channel decomposition of $I_d$ by computing it separately on the plaquette channel, the Higgs channel, and their union. Two diagonal scans in the $(\beta,\beta_H)$ plane, each crossing two transition lines of different character, then allow us to compare how each channel responds to transitions driven by different degrees of freedom, and whether the response of $I_d$ depends on which physical sector is included in the distance metric.
 
Together these benchmarks address the two capabilities outlined above. Beyond QCD multi-transition settings, the same geometric diagnostics may also find use in other complex systems that host several phases within parameter space.

The remainder of this paper is organized as follows. \cref{sec: method} reviews the 2NN intrinsic dimension estimator and its physical relation to correlations. \cref{sec: Results} presents the lattice setups and numerical results, including clock-model scans for $q=5, 9$ and $U(1)$ Higgs scans along two diagonal trajectories in the $(\beta,\beta_H)$ plane, each compared with conventional thermodynamic observables. \cref{sec: discussion} discusses the implications of these benchmark results for possible future applications to non-Abelian gauge theories and full QCD, including the construction of physically meaningful gluonic and fermionic channels. In \cref{sec: conclusion} we summarize the work.

\section{Intrinsic Dimension Estimation}\label{sec: method}

The Two-Nearest-Neighbors (2NN) method is a robust, non-parametric approach for estimating the intrinsic dimensionality of a dataset embedded in a high-dimensional space~\cite{Facco:2017est}. By focusing exclusively on the distances to the first two nearest neighbors, the method restricts the assumption of constant density to a minimal local neighborhood, thereby strongly reducing sensitivity to global density variations and local geometric distortions.

\subsection{Theoretical Foundation}

Consider a dataset $\mathcal{D} = \{\mathbf{x}_1, \ldots, \mathbf{x}_N\}$ sampled from a manifold of intrinsic dimension $d$, embedded in an ambient space $\mathbb{R}^D$ ($d \le D$). Around any reference point $\mathbf{x}_i$, we assume the points are locally distributed according to a Poisson process with a uniform local density $\rho$.

The volume of a $d$-dimensional ball of radius $r$ is $V(r) = \omega_d r^d$, where $\omega_d$ is the volume of the unit $d$-ball. According to Poisson statistics, the probability of finding no points within a radius $r_1$ is governed by
\begin{equation}
    P(k=0 | r_1) = \exp\left(-\rho \omega_d r_1^d\right).
    \label{eq:poisson_void}
\end{equation}
The probability density function (PDF) for the distance to the first nearest neighbor, $r_1$, is then derived by multiplying this void probability by the expected number of points in the infinitesimal shell $dr_1$:
\begin{equation}
    f_{1}(r_1) = \rho d \omega_d r_1^{d-1} \exp\left(-\rho \omega_d r_1^d\right).
    \label{eq:pdf_r1}
\end{equation}

Similarly, the conditional probability of finding the second nearest neighbor at a distance $r_2$ ($r_2 > r_1$), given that the shell $(r_1, r_2)$ is empty, allows us to construct the joint probability density $f(r_1, r_2) = f_1(r_1) f(r_2 | r_1)$, which yields
\begin{equation}
    f(r_1, r_2) = (\rho d \omega_d)^2 r_1^{d-1} r_2^{d-1} \exp\left(-\rho \omega_d r_2^d\right).
\end{equation}

The key insight of the 2NN method lies in defining the dimensionless ratio $\mu = r_2/r_1$ ($\mu \ge 1$). By performing a change of variables from $(r_1, r_2)$ to $(r_1, \mu)$ and integrating out the scale-dependent variable $r_1$, the dependence on the local density $\rho$ completely cancels out. The marginal PDF for $\mu$ rigorously evaluates to a Pareto distribution:
\begin{equation}
    f(\mu) = d \, \mu^{-(d+1)}, \quad \mu \geq 1.
    \label{eq:pareto}
\end{equation}
The independence of $f(\mu)$ from $\rho$ implies that the distribution of $\mu$ depends only on the intrinsic dimension $d$, making the estimator robust to slowly varying sampling densities.

\subsection{Practical Estimation of Intrinsic Dimension}

To estimate the theoretical intrinsic dimension (which we denote as $I_d$ in practice) for the dataset $\mathcal{D}$, we compute the ratio $\mu_i = r_{i,2}/r_{i,1}$ for each point $\mathbf{x}_i$. Following the distribution derived in \cref{eq:pareto}, the PDF parameterized by the intrinsic dimension estimator $I_d$ is
\begin{equation}
    f(\mu) = I_d \mu^{-(I_d+1)}, \quad \mu \in [1, \infty).
    \label{eq:2nn_pdf}
\end{equation}

The corresponding theoretical cumulative distribution function (CDF) is given by
\begin{equation}
    F(\mu) = \int_1^\mu f(x) \, dx = 1 - \mu^{-I_d}.
    \label{eq:2nn_cdf}
\end{equation}

By equating $F(\mu)$ to the empirical CDF constructed from the data, $F_{\text{emp}}(\mu)$, we obtain the linear scaling relation:
\begin{equation}
    \ln \left[ 1 - F_{\text{emp}}(\mu) \right] = -I_d \ln(\mu).
    \label{eq:linear_relation}
\end{equation}
This allows $I_d$ to be straightforwardly estimated via linear regression passing through the origin.

\subsection{Physical Interpretation}

Applying the 2NN estimator to an ensemble of Monte Carlo (MC) configurations requires some care, since MC simulations sample a globally nonuniform state space according to the Boltzmann distribution, $P \propto \exp(-\beta H)$. The 2NN derivation, however, requires approximate uniformity only locally, over the scale set by the first two nearest neighbors. Two conditions are particularly relevant in practice. First, the sampled
configurations should be sufficiently decorrelated; this can be achieved by saving configurations at intervals larger than the integrated autocorrelation
time. Second, the sampling density should vary slowly within the neighborhood of radius $r_2$. This condition is plausible when neighboring configurations
have similar Boltzmann weights, but it is not guaranteed \textit{a priori} and should be checked empirically. In addition, exact distance degeneracies
($r_1=r_2$) should be negligible, since they are incompatible with the continuous Pareto distribution assumed by the estimator.

Provided these conditions are met, the quantity $I_d$ estimated from the MC configurations can be regarded as a geometric property of the sampled configuration space. Rather than being derived from a preselected operator, $I_d$ measures the effective number of independent degrees of freedom needed to describe the ensemble locally. It is therefore sensitive to how correlations reorganize the sampled measure as a control parameter is varied.

In real space, two-point correlation functions of chosen operators define a correlation length $\xi$. When $\xi$ becomes large compared with the microscopic scales of the lattice, long-range correlations reduce the number of effectively independent field variables, and the geometry of the sampled configuration space changes. Correspondingly, $I_d$ develops a characteristic feature near the transition, which may appear as a minimum, a plateau, or a change of slope. For continuous second-order transitions, finite-size analyses of classical and quantum many-body systems have shown that $I_d$ admits a scaling collapse of the form $I_d = L^{\zeta} f(\xi/L)$, with $\xi \sim |T-T_c|^{-\nu}$ and $\zeta$ a nonuniversal scaling exponent associated with $I_d$~\cite{Mendes-Santos:2020msa,Mendes-Santos:2021mhs,Panda:2023zcj}. Analogous geometric signatures are also found at Berezinskii--Kosterlitz--Thouless (BKT) transitions, where $\xi$ diverges exponentially rather than as a power law~\cite{Mendes-Santos:2020msa}.

This perspective clarifies how $I_d$ complements conventional thermodynamic probes. A susceptibility $\chi$ peaks where the variance of one selected observable is maximal. By contrast, $I_d$ responds to changes in the joint distribution of the field variables that enter the distance measure. The two diagnostics therefore need not locate a crossover at exactly the same coupling, even when they identify the same transition. There is in general no closed-form map from a two-point correlator $C(r)$ to $I_d$; the utility of the estimator is that long-range correlations leave a measurable signature in the geometry of the sampled configuration space, which can be extracted without specifying an order parameter \textit{a priori}.

\section{Numerical Results}\label{sec: Results}

A Monte Carlo configuration $C$ comprises the full set of field variables on the lattice. For a single-field model such as the clock model, $C$ is one array. For a multi-field model such as the $U(1)$ Higgs model, $C$ decomposes naturally into sectors with distinct degrees of freedom. We treat each sector as an independent channel with its own distance metric, and evaluate $I_d$ either on a single channel or on their union.

At each temperature or coupling we generate $N_{\rm cfg}=5\times10^{4}$ equilibrated configurations, used both for thermodynamic observables and for the intrinsic-dimension analysis. For all ensembles, sufficient thermalization is performed before measurements, and configurations are retained at a fixed separation much larger than the integrated autocorrelation time $\tau_{\rm int}$ of bulk observables, estimated from short dedicated runs at representative parameter values. Model-specific details are given below. The retained configurations can thus be regarded as effectively independent, as required by the local-uniformity assumption underlying the 2NN estimator. A consistency check for the clock model is presented in Appendix~\ref{app:linear}. For $I_d$, this sample is partitioned into five statistically independent subsets of $10^{4}$ configurations each. We report the mean and error bar obtained from the five independent 2NN estimates.

\subsection{$q$-State Clock Model}\label{sec: results_clock}

The $q$-state clock model is defined on a two-dimensional square lattice with linear size $L$ and periodic boundary conditions~\cite{Jose:1977gm,Potts:1951rk}. Each lattice site $i$ carries a planar spin variable $\boldsymbol{S}_i$ that is constrained to point along one of $q$ equally-spaced directions,
\begin{equation}
\boldsymbol{S}_i = (\cos\theta_i, \sin\theta_i), \quad \theta_i = \frac{2\pi n_i}{q}, \quad n_i \in \{0, 1, 2, \ldots, q-1\}.
\end{equation}
The Hamiltonian of the system is
\begin{equation}
H = -J \sum_{\langle i,j \rangle} \boldsymbol{S}_i \cdot \boldsymbol{S}_j = -J \sum_{\langle i,j \rangle} \cos(\theta_i - \theta_j),
\label{eq:clock_hamiltonian}
\end{equation}
where $J > 0$ is the ferromagnetic coupling and the sum runs over all nearest-neighbor pairs $\langle i,j \rangle$. We set $J = 1$ throughout this work to define the energy scale. The partition function at inverse temperature $\beta = 1/T$ reads
\begin{equation}
Z = \sum_{\{\theta_i\}} \exp\left(\beta J \sum_{\langle i,j \rangle} \cos(\theta_i - \theta_j)\right),
\end{equation}
where the sum is taken over all possible spin configurations. Equilibrium configurations are generated using the single-spin-flip Metropolis algorithm. For each lattice size and temperature, we discard $1.8\times10^{4}$ sweeps from a random initial configuration and subsequently retain one configuration every $1.3$--$1.5\times10^{4}$ sweeps, with the larger separation used in the transition region. Five statistically independent runs, with different random seeds and initial configurations, yield $10^{4}$ configurations each, for a total of $N_{\rm cfg}=5\times10^{4}$ per temperature.

A configuration is stored as an integer array, with each site taking values $n_i\in\{0,1,\ldots,q-1\}$. To compute the distance between two configurations $\mathbf{x}^{(i)}$ and $\mathbf{x}^{(j)}$, we map the discrete states onto the unit circle to respect the periodic topology of the angular variables:
\begin{equation}
d(\mathbf{x}^{(i)}, \mathbf{x}^{(j)}) = \sqrt{\sum_{k=1}^{N}\left[1 - \cos(\theta_k^{(i)} - \theta_k^{(j)})\right]}.
\end{equation}
Crucially for the 2NN method, while the individual spin variables are highly discrete, tracing this Euclidean distance in the embedding space over a large number of lattice sites $N$ generates a quasi-continuous distance spectrum. This effectively bypasses exact distance degeneracies and satisfies the continuity requirement of the Pareto distribution, as confirmed by the strict linear collapse through the origin of the empirical CDF in Fig.~\ref{fig:2nn_example} (an exact degeneracy $r_1=r_2$ would instead produce a nonzero intercept at $\ln\mu=0$).

The $q$-state clock model provides a useful benchmark because the separation between its two BKT transitions can be varied by changing $q$~\cite{Jose:1977gm,Li:2019dkb,Goswami:2025ptc}. We consider $q$=9, for which the two transitions are well separated, and $q$=5, where they occur in a much narrower temperature interval. Conventional thermodynamic observables are also computed to provide an independent characterization of the transitions. The average energy density $E$, magnetization $M$, and magnetic susceptibility $\chi$ are defined as
\begin{align}
E &= \frac{1}{N}\langle H \rangle, \quad 
M = \frac{1}{N}\left\langle \left|\sum_i \mathbf{S}_i\right|\right\rangle, \nonumber\\
\chi &= \frac{\beta}{N}\left(\langle M_{\rm tot}^2\rangle - \langle M_{\rm tot}\rangle^2\right),
\end{align}
where $M_{\rm tot} = \left|\sum_i \mathbf{S}_i\right|$, and $N=L^2$.

\subsubsection{$q=9$ Clock Model}

\begin{figure}[htbp!]
    \centering
    \includegraphics[width=0.5\textwidth]{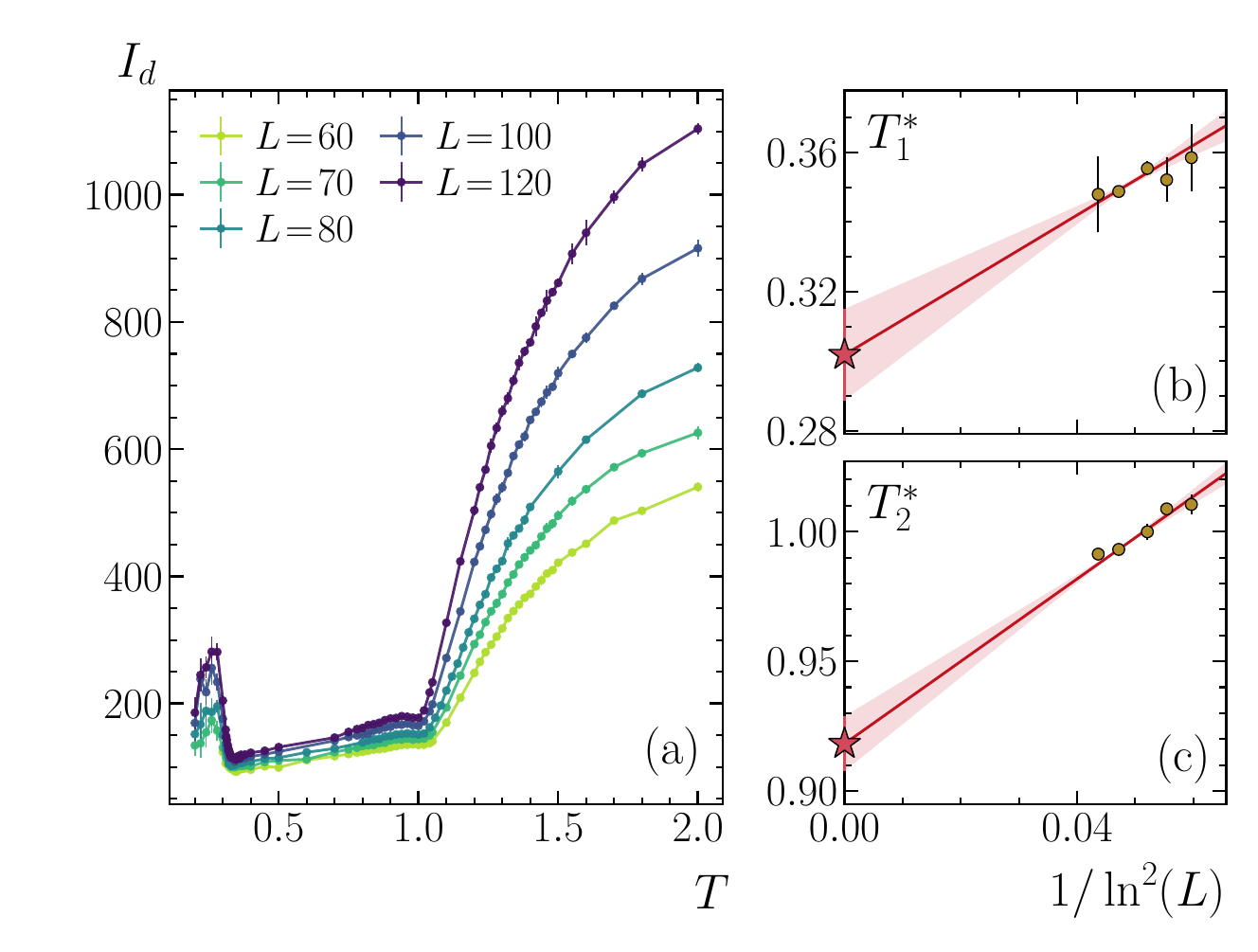}
    \caption{$q=9$ clock model. Left: temperature dependence of $I_d$ for different values of $L$ ($10^{4}$ configurations per run, with means and error bars from five independent runs). The pseudo-critical temperature $T^*(L)$ is identified as the location of the minimum of $I_d(T)$ for each lattice size $L$.} Right: finite-size scaling of the pseudo-critical temperatures $T_1^*$ (top) and $T_2^*$ (bottom) versus $1/\ln^2 L$. Error bars on $T^{*}$ are bootstrap estimates, and the uncertainty of the intercept is propagated from the fit covariance. Linear extrapolations yield $T_1(\infty)=0.302(13)$ and $T_2(\infty)=0.918(10)$.
    \label{fig:NineId}
\end{figure}

\begin{figure*}[htbp!]
    \centering
    \includegraphics[width=\textwidth]{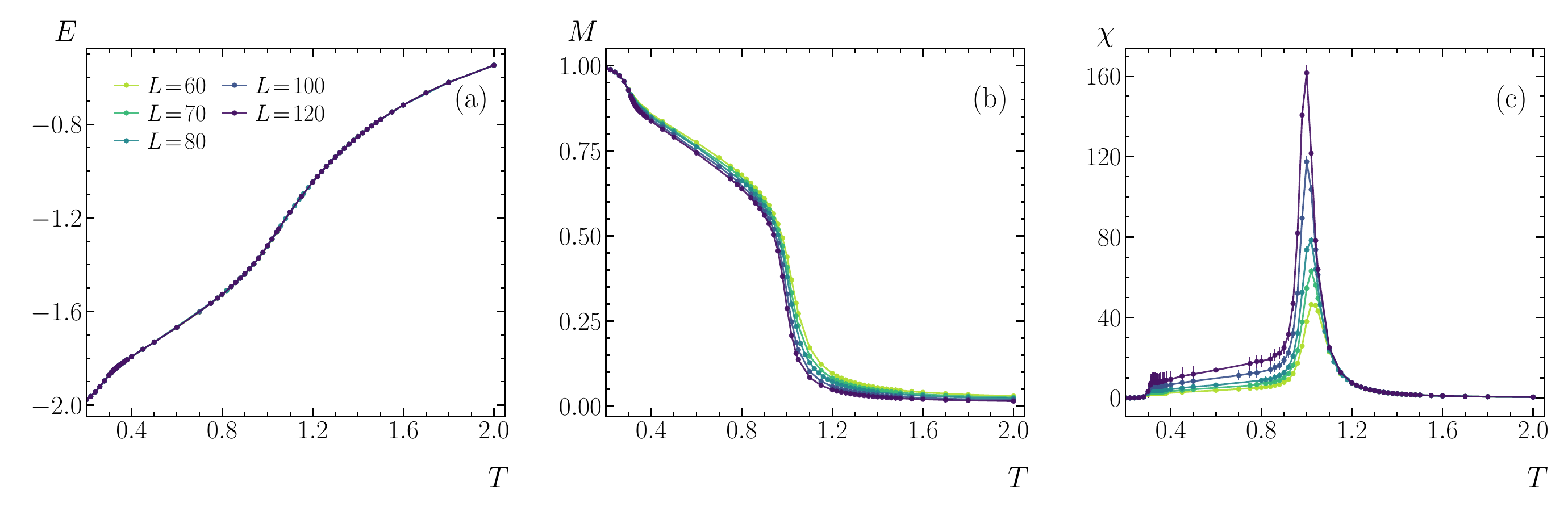}
    \caption{$q=9$ clock model. Temperature dependence of the average energy $E$, magnetization $M$, and magnetic susceptibility $\chi$ for different values of $L$. The data set comprises $N_{\rm cfg}=5\times 10^{4}$ configurations.}
    \label{fig:NineThermo}
\end{figure*}

\cref{fig:NineId} shows the temperature dependence of $I_d$ 
alongside the finite-size scaling used to extract the critical temperatures. The evolution of $I_d$ captures three distinct phases of the model. The minimum of $I_d(T)$ near each transition provides a pseudo-critical temperature $T^{*}(L)$, which we locate by a weighted quartic fit of $I_d$
over a temperature window bracketing each of the minimum, with the uncertainty of $T^{*}$ obtained from a parametric bootstrap over the error bars of $I_d$, as discussed in~\cite{Mendes-Santos:2020msa}. In the low-temperature ordered phase ($T \lesssim 0.34$) for the $L$ considered, the spontaneous breaking of the $Z_9$ symmetry highly constrains the configuration space. $I_d$ exhibits a sharp initial peak before dropping to a local minimum near $T_1$.

Entering the quasi-critical (BKT) phase ($0.34 \lesssim T \lesssim 0.92$), the system is characterized by emergent $U(1)$ symmetry and bound vortex-antivortex pairs. Here, $I_d$ exhibits a broad, relatively flat basin with only a slow, stable rise ($I_d \sim 100-150$). This stability is a geometric signature of the scale-invariant, algebraically decaying correlations, which globally suppress the effective degrees of freedom despite the increasing temperature. However, as the temperature crosses the upper BKT transition ($T \gtrsim 0.92$) into the disordered phase, the vortex-antivortex pairs unbind, and the correlation length becomes finite, decreasing rapidly as the system moves away from the transition.

A rigorous determination of $T_1$ and $T_2$ is achieved via the finite-size scaling of the pseudo-critical temperatures $T^*_{1,2}$. Since both transitions belong to the BKT universality class, they obey the asymptotic scaling ansatz
\begin{align}
    T^*(L) - T_{\rm BKT} \sim \frac{1}{\ln^2 L}.
    \label{eq:FSS_BKT}
\end{align}
 \cref{eq:FSS_BKT} follows from the general argument that the pseudo-critical crossover occurs when the correlation length $\xi(T)$ reaches the system size $L$. Substituting the BKT form $\xi\sim e^{a/(T-T_{\rm BKT})^{1/2}}$ and solving for $T^*(L)$ gives the $1/\ln^2 L$ scaling (see, e.g.,~\cite{Sandvik:2010zza}). We fit $T^*(L) = T_{\rm BKT} + a'/\ln^2(L)$ with $T_{\rm BKT}$ and $a'$ as free parameters, shown as the linear extrapolations in \cref{fig:NineId} (right).

The extrapolations shown in the right panels of \cref{fig:NineId} yield $T_1=0.302(13)$ and $T_2=0.918(10)$ for $q=9$. These values align well with theoretical expectations; particularly, $T_2$ is fundamentally consistent with the limiting XY model ($q\rightarrow\infty$) BKT temperature $T_{\rm BKT} \approx 0.893$, slightly modified by finite-$q$ relevant perturbations.

\begin{figure}[htbp!]
    \centering
\includegraphics[width=0.5\textwidth]{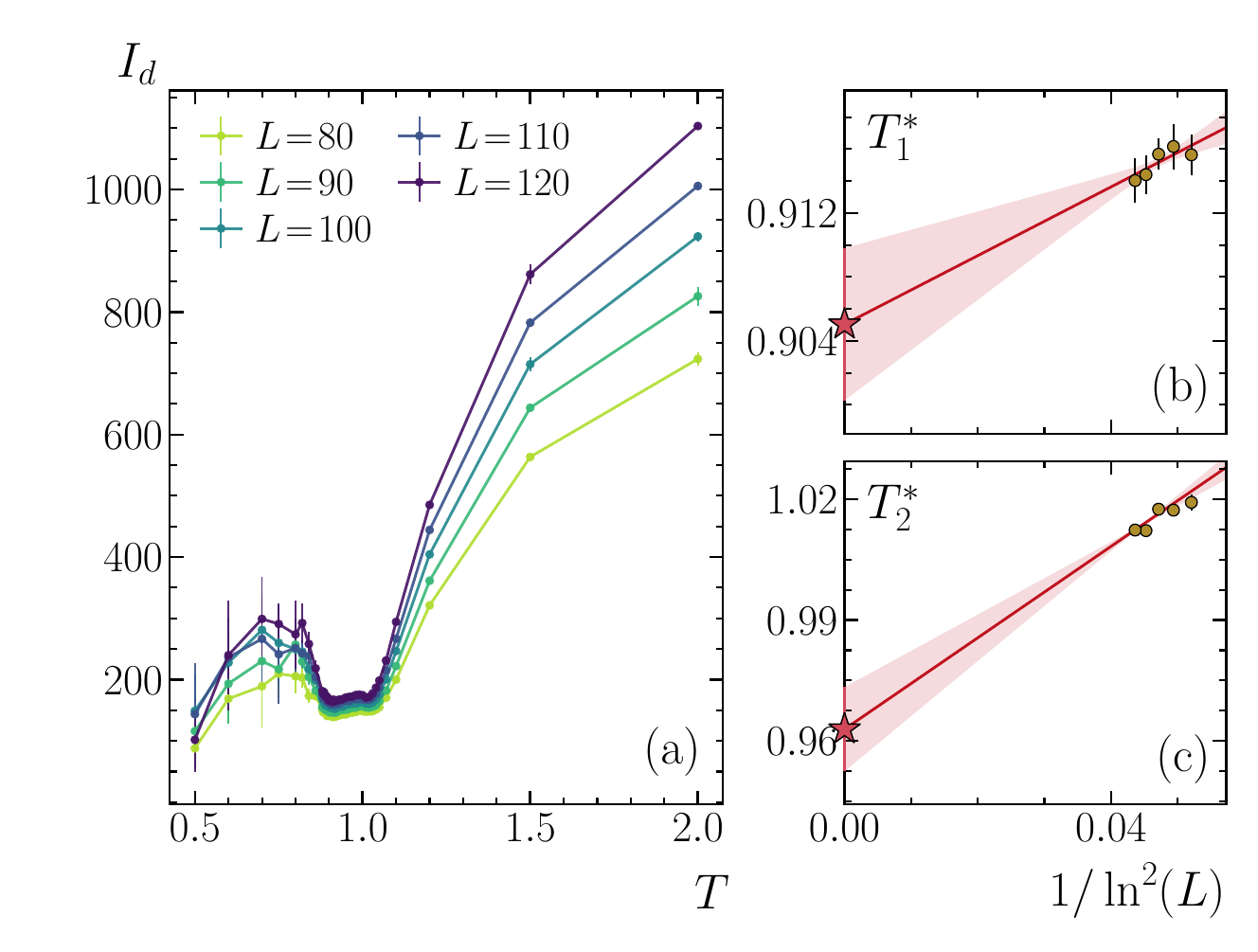}
    \caption{$q=5$ clock model. Left: temperature dependence of $I_d$ for different values of $L$ ($10^{4}$ configurations per run, with means and error bars from five independent runs). The pseudo-critical temperature $T^*(L)$ is identified as the location of the minimum of $I_d(T)$ for each lattice size $L$. 
    Right: finite-size scaling of $T_1^*$ (top) and $T_2^*$ (bottom) versus $1/\ln^2 L$. Error bars on $T^{*}$ are bootstrap estimates, and the
    uncertainty of the intercept is propagated from the fit covariance. Linear extrapolations yield $T_1(\infty)=0.905(05)$ and $T_2(\infty)=0.963(10)$.}
    \label{fig:FiveId}
\end{figure}

\begin{figure*}[htbp!]
    \centering
    \includegraphics[width=\textwidth]{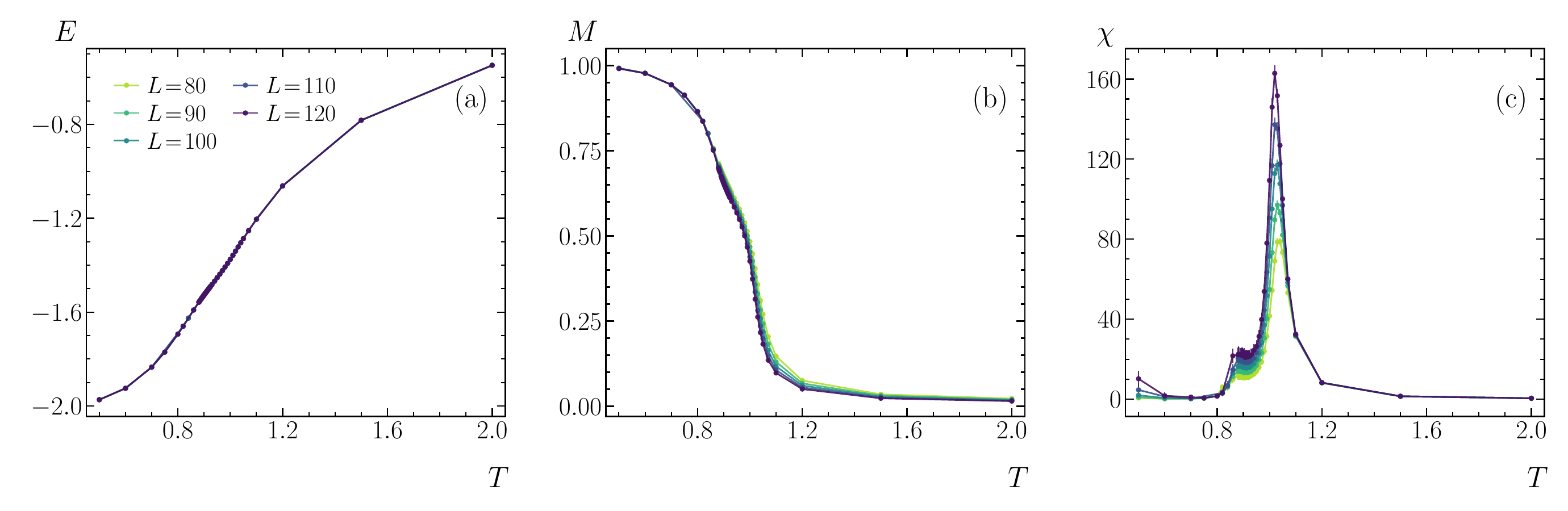}
    \caption{$q=5$ clock model. Temperature dependence of the average energy $E$, magnetization $M$, and magnetic susceptibility $\chi$ for different values of $L$. The data set comprises $N_{\rm cfg}=5\times 10^{4}$ configurations.}
    \label{fig:FiveThermo}
\end{figure*}

The conventional observables(\cref{fig:NineThermo}) further validate this picture. The energy $E$ evolves smoothly without latent heat, confirming the continuous nature of the transitions. The magnetization $M$ is saturated in the low-temperature phase and only decays fully to zero above $T_2 \approx 0.92$, precisely where $I_d$ exhibits its steep upsurge. The magnetic susceptibility $\chi$ presents a massive divergence at the upper transition $T_2$ alongside a much weaker shoulder near $T_1$, unambiguously framing the two separated transitions and setting the stage for the $q=5$ case.

\subsubsection{$q=5$ Clock Model}\label{sec: results_q5}

The temperature dependence of $I_d$ and its finite-size scaling for the $q=5$ clock model are presented in \cref{fig:FiveId}. This model exhibits a more subtle phase structure, with two BKT transitions occurring within a narrow temperature range~\cite{Li:2019dkb,Surungan:2019hjf}. The close proximity of the two transition temperatures makes them difficult to resolve using conventional observables.

Notably, the intrinsic dimension provides a natural way to resolve this complex phase structure. After an initial low-temperature peak around $T \approx 0.7$, $I_d$ drops and carves out a pronounced, stable minimum (a flat valley) extending roughly from $T \approx 0.85$ to $1.0$. The finite-size scaling analysis (\cref{fig:FiveId}, right panels) locates the two transitions at $T_1=0.905(05)$ and $T_2=0.963(10)$ (in agreement with established literature values $T_1=0.9059$ and $T_2=0.9521$~\cite{Li:2019dkb}). 
This broad geometric valley in $I_d$ 
captures the very narrow quasi-critical intermediate phase ($0.90 \lesssim T \lesssim 0.96$). Beyond $T \approx 1.0$, $I_d$ resumes its aggressive ascent as the system melts into the fully disordered phase, restoring the strong $L$-dependence seen in the $q=9$ case.

The usefulness of $I_d$ is most clearly seen by comparing it with the conventional observables in \cref{fig:FiveThermo}.
Because $T_1$ and $T_2$ are very close, the energy and magnetization show only smooth crossovers and do not clearly resolve the two transitions. The susceptibility $\chi$, in contrast, exhibits a closely spaced double-peak structure and thus identifies the two phase boundaries. However, $\chi$ mainly signals the transition points themselves, whereas $I_d$ provides complementary information about the structure of the intermediate phase. As shown in \cref{fig:FiveId}, $I_d$ forms a stable, non-extensive valley throughout the interval between $T_1$ and $T_2$, rather than showing sharp fluctuations only at the boundaries. This behavior directly reveals that the configuration-space manifold retains a low-dimensional, scale-invariant structure throughout the quasi-critical phase.

\subsection{$U(1)$ Higgs Model}\label{sec: method_u1higgs}

The four-dimensional $U(1)$ Higgs model is defined on a hypercubic lattice of linear size $L$ with periodic boundary conditions~\cite{Callaway:1981rt,Montvay:1994cy}. The degrees of freedom consist of compact gauge link variables $U_{n,\mu}\in U(1)$ residing on the links and a charged scalar (Higgs) field $\varphi_n$ residing on the sites. Throughout this work we adopt the London (fixed-length) limit $|\varphi_n|=1$, so that $\varphi_n$ is a pure $U(1)$ phase, as in the classical Monte Carlo surveys of the model~\cite{Callaway:1981rt,Bowler:1981cj}. The action takes the form
\begin{equation}
A = \beta\sum_{\substack{n,\mu,\nu\\ \mu<\nu}} A_{\square}(n,\mu,\nu)
  + \beta_H\sum_{n,\mu}A_H(n,\mu),
  \label{eq: action}
\end{equation}
where the plaquette action $A_\square$ and the scalar action $A_H$ are given by
\begin{align}
A_\square(n,\mu,\nu) &= 1 - \mathrm{Re}\,
  \bigl(U_{n,\mu}U_{n+\hat\mu,\nu}U_{n+\hat\nu,\mu}^{-1}U_{n,\nu}^{-1}\bigr),\\
A_H(n,\mu) &= \frac{1}{2}\left| \varphi_{n+\hat\mu} - U^{Q}_{n,\mu}\varphi_n \right|^{2}.
\end{align}
Here $\hat{\mu}$ and $\hat{\nu}$ denote unit vectors along the respective lattice directions, and $Q$ is the $U(1)$ charge of the scalar field. With the conventional identification $U_{n,\mu}=e^{i\theta_{n,\mu}}$ and $\varphi_n=e^{i\varphi_n}$, the plaquette term reduces to $A_\square=1-\cos\theta_\square$ with $\theta_\square$ the plaquette angle, and the hopping term reduces to
$A_H = 1-\cos\!\left(\varphi_{n+\hat\mu}-\varphi_n-Q\,\theta_{n,\mu}\right)$. The coupling $\beta$ controls the gauge sector, while $\beta_H$, the hopping parameter of the scalar field, controls the Higgs sector. We set $Q=2$ throughout.

\begin{figure}[htbp!]
    \centering
    \includegraphics[width=0.48\textwidth]{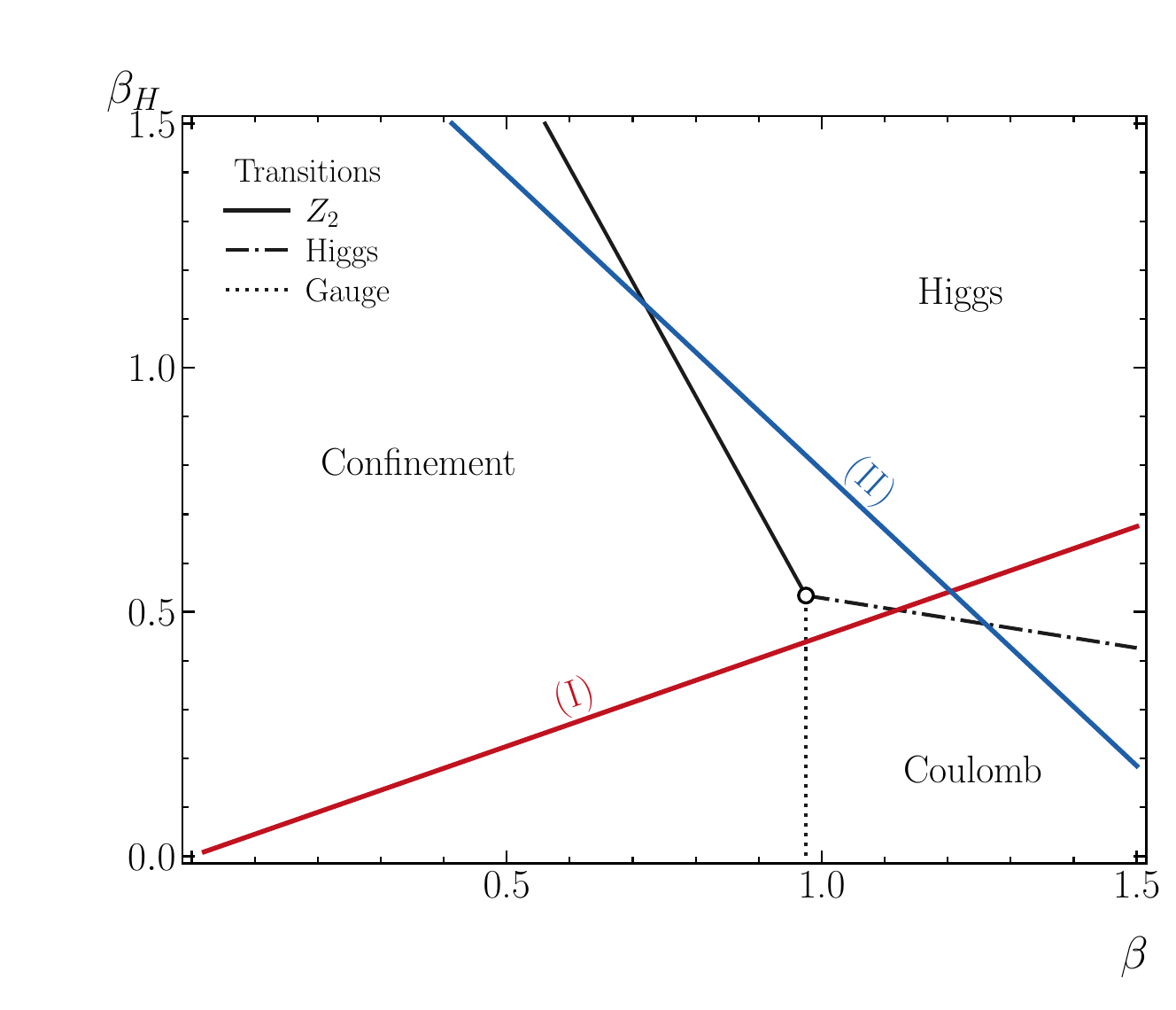}
    \caption{Schematic phase diagram of the four-dimensional $U(1)$ Higgs model in the $(\beta,\beta_H)$ plane~\cite{Callaway:1981rt}, showing the confinement, Coulomb, and Higgs regions separated by the gauge, Higgs, and $Z_2$ transition lines. The two diagonal scan trajectories used in this work are overlaid: line (I) $\beta_H=0.45\,\beta$ (red) and line (II) $\beta_H=-1.2\,\beta+2$ (blue).}
    \label{fig:u1higgs_phase}
\end{figure}

The phase diagram of this model, illustrated in \cref{fig:u1higgs_phase}, has been studied extensively and exhibits three regions separated by the gauge, Higgs, and $Z_2$ transition lines~\cite{Fradkin:1978dv,Callaway:1981rt,Bowler:1981cj,Jansen:1985nh}. In the strong-coupling regime (small $\beta$, small $\beta_H$), the system resides in the confinement phase, dominated by non-linear gauge fluctuations and the proliferation of compact-$U(1)$ monopoles~\cite{Polyakov:1976fu}. This produces a highly disordered configuration space in which fundamental test charges are linearly confined. Increasing $\beta$ at small $\beta_H$ drives a gauge transition into the Coulomb phase, a gapless regime with a massless photon whose existence in four-dimensional compact $U(1)$ is rigorously established~\cite{Guth:1979gz}; this phase occupies the large-$\beta$, small-$\beta_H$ region. 
At large $\beta_H$ and sufficiently large $\beta$, 
the system enters the Higgs phase. Because the scalar carries charge $Q=2$ rather than the unit charge ($Q=1$), the Higgs mechanism leaves a residual $Z_2$ gauge structure, allowing a genuine boundary between the confinement and Higgs regions~\cite{Fradkin:1978dv,Banks:1979fi}. This is in contrast to the $Q=1$ case where the Higgs and confinement regions are analytically connected. Across this line, gauge and matter degrees of freedom reorganize simultaneously. The Higgs and Coulomb phases are, in turn, separated by the Higgs transition line at large $\beta$.

In this work, we follow two diagonal trajectories in the $(\beta,\beta_H)$ plane, each designed to cross two transition lines of distinct physical character (see \cref{fig:u1higgs_phase}):
\begin{align}
\text{(I)}\quad \beta_H &= 0.45\,\beta,\nonumber\\
\text{(II)}\quad \beta_H &= -1.2\,\beta + 2.
\label{eq: u1higgs_lines}
\end{align}
Trajectory (I) crosses the confinement--Coulomb (gauge) and Coulomb--Higgs (Higgs) boundaries; trajectory (II) crosses the confinement--Higgs ($Z_2$) and Higgs--Coulomb (Higgs) boundaries. Equilibrium configurations are generated using a combined heat-bath and overrelaxation update of both the gauge links and the Higgs field. One update cycle consists of a heat-bath sweep over the Higgs phases, a heat-bath sweep over the gauge links, and $n_{\rm or}=4$ overrelaxation cycles, each comprising one reflection sweep of the Higgs phases and one of the links. Each sweep updates the even and odd sublattices in turn. Since the two terms in \cref{eq: action} do not yield a common local effective field for a link variable, the heat-bath proposal for each link is constructed from the plaquette staple alone, while the scalar hopping term is incorporated through a Metropolis accept-reject step. The same correction is applied to the overrelaxation reflections to ensure detailed balance with respect to the full action.

At each coupling point, we discard $2\times10^{3}$--$3\times10^{3}$ update cycles for thermalization and retain one configuration every $40$--$150$ cycles, with larger separations used for larger lattices. Away from the $Z_2$ transition, the lag-one autocorrelation coefficients of $P$ and $H$ remain below $0.02$ at all couplings, indicating that the retained configurations are effectively independent. To improve sampling near the first-order $Z_2$ boundary, we additionally perform replica exchange between neighboring coupling points along the trajectory, with exchanges attempted every $10$ update cycles. Exchanges are restricted to chains with the same replica index, so that the five subsets used to estimate the mean and error bar of $I_d$ remain mutually independent.

\begin{figure}[htbp!]
    \centering
    \includegraphics[width=\columnwidth]{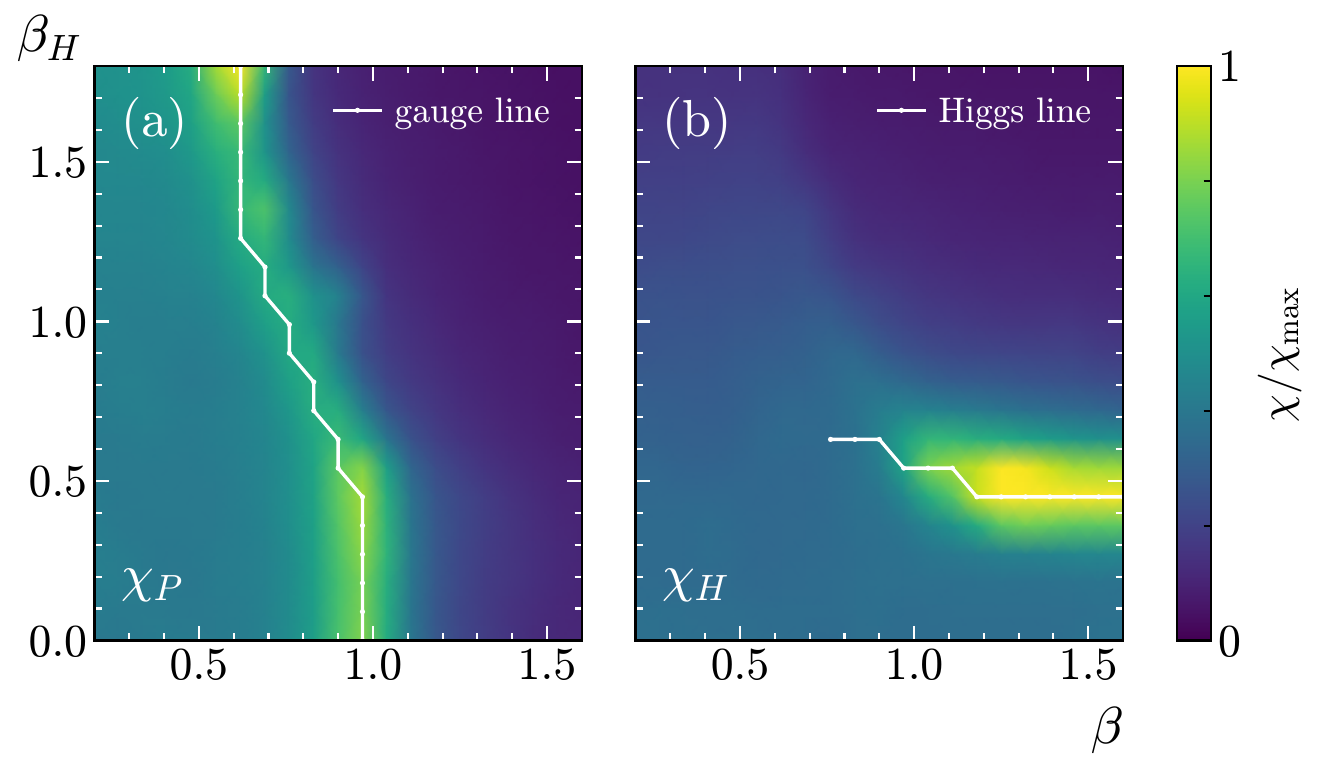}
    \caption{Susceptibility heatmaps over the $(\beta,\beta_H)$ plane at $L=8$, each normalised to its own maximum: (a) the plaquette susceptibility $\chi_{_P}$ (gauge channel) and (b) the Higgs susceptibility $\chi_{_H}$ (Higgs channel). In each panel the overlaid curve traces the ridge of that panel's own susceptibility.}
    \label{fig:heatmaps}
\end{figure}

\begin{figure*}[htbp!]
    \centering
    \includegraphics[width=\textwidth]{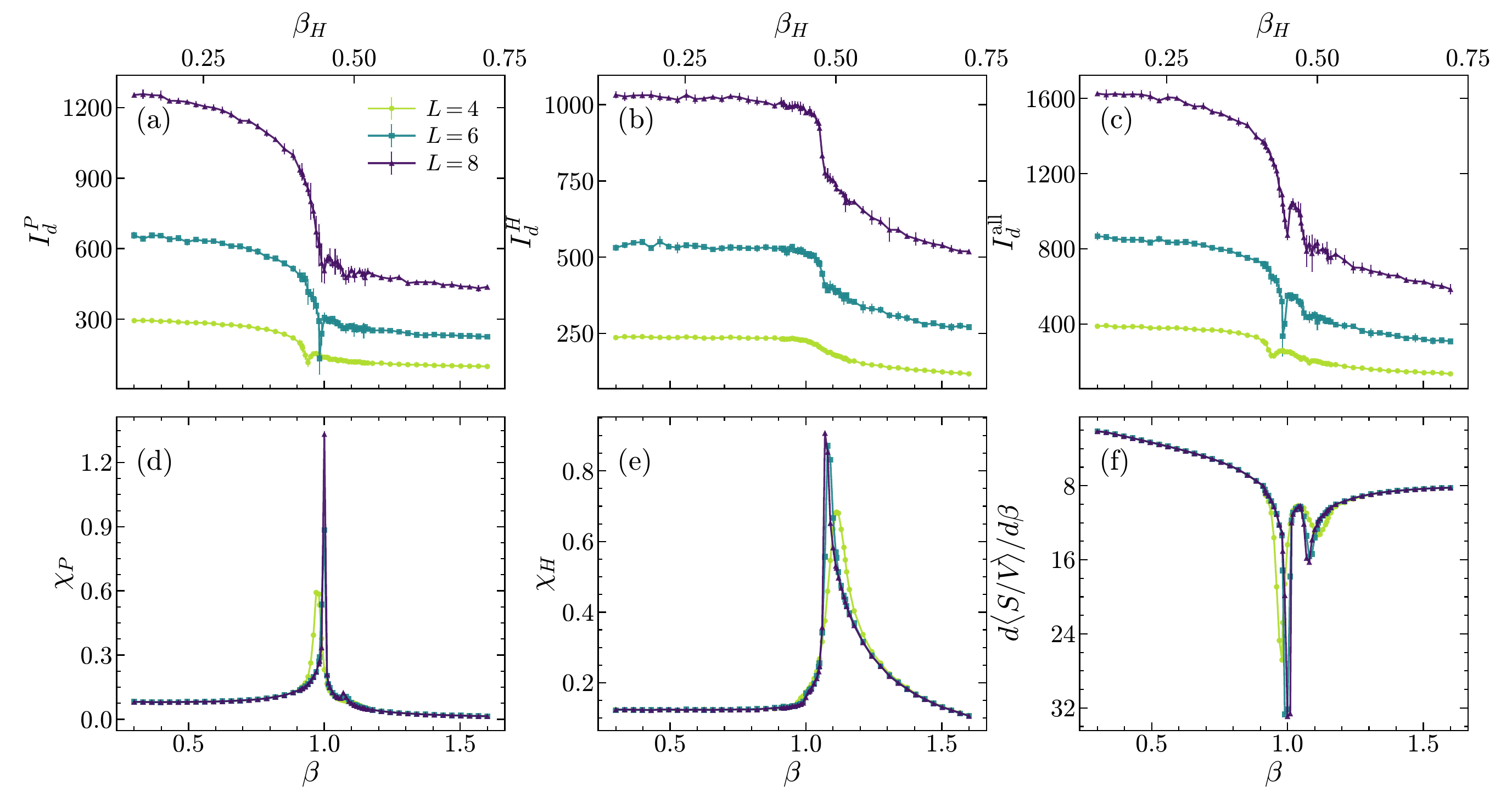}
    \caption{$U(1)$ Higgs model along trajectory (I), $\beta_H=0.45\,\beta$, for $L=4,6,8$. Top row: intrinsic dimension evaluated on (a) the plaquette channel $I_d^P$, (b) the Higgs channel $I_d^H$, and (c) the combined channel $I_d^{\rm all}$. Bottom row: (d) plaquette susceptibility $\chi_P$, (e) Higgs susceptibility $\chi_H$, and (f) the derivative of the mean action density $d\langle S/V\rangle/d\beta$. The bottom axis shows $\beta$ and the top axis the corresponding $\beta_H$.}
    \label{fig:u1higgs_line1}
\end{figure*}

A configuration $C$ in this model is fully specified by the gauge link angles $\{\theta_{n,\mu}\}$ and the Higgs phases $\{\varphi_n\}$. To probe how each sector responds to the transitions encountered along each trajectory, we introduce a channel decomposition of the configuration space.\footnote{Before introducing physical channels, we verified that raw gauge-variant variables fail because nearest-neighbor geometry is dominated by gauge orbits; see Appendix~\ref{app:gauge_redundancy}.}
The \textit{plaquette channel} is constructed from the six gauge-invariant plaquette orientations per lattice site,
\begin{equation}
P_{\mu\nu}(n) = \bigl[\theta_{n,\mu} + \theta_{n+\hat\mu,\nu}
              - \theta_{n+\hat\nu,\mu} - \theta_{n,\nu}\bigr]\bmod 2\pi,
\quad \mu < \nu,
\end{equation}
which encode the gauge sector. The \textit{Higgs channel} is constructed from the four gauge-invariant covariant phase differences
\begin{equation}
H_\mu(n) = \bigl[\varphi_{n+\hat\mu} - \varphi_n - Q\,\theta_{n,\mu}\bigr]\bmod 2\pi,
\end{equation}
which encode the matter sector. We additionally define a \textit{combined channel} formed by the union of both variable sets to probe the global configuration. Since all channel variables are angular quantities on $[0,2\pi)$, the geometric distance between two configurations $C$ and $C'$ is naturally defined via the Euclidean embedding of the angles:
\begin{equation}
d(C,C') = \sqrt{\,\sum_i \left[1 - \cos\!\left(v_i(C)-v_i(C')\right)\right]\,},
\label{eq: u1higgs_distance}
\end{equation}
where $\{v_i\}$ denotes the relevant set of channel variables. By evaluating $I_d$ independently on each channel along a single trajectory, we can isolate which sector of degrees of freedom carries the primary signal of each transition, thereby revealing the selectivity of the intrinsic dimension to underlying physical mechanisms.

Apart from the intrinsic dimension, we also calculate conventional observables for comparison, which are given by
\begin{align}
  P&=\frac{1}{6V}\sum_{n,\mu<\nu}\cos\theta_\square(n,\mu,\nu),\qquad
  H=\frac{1}{4V}\sum_{n,\mu}\cos H_\mu(n), \nonumber\\
  \chi_P&=V\big(\langle P^2\rangle-\langle P\rangle^2\big),\qquad
  \chi_H=V\big(\langle H^2\rangle-\langle H\rangle^2\big), \nonumber\\
  \frac{S}{V}&=-\big(6\beta P+4\beta_H H\big),
\end{align}
with $V=L^4$. Note that $P$ and $H$ are normalized per plaquette and per link
while both susceptibilities carry a single factor $V$.

To set the stage for the trajectory scans, we first visualize the global phase structure using conventional thermodynamic susceptibilities. \cref{fig:heatmaps} shows the plaquette susceptibility $\chi_P$ and the Higgs susceptibility $\chi_H$ computed over the full $(\beta,\beta_H)$ plane at $L=8$. The ``gauge line'' tracks the $\chi_P$ ridge, while the ``Higgs line'' tracks the $\chi_H$ ridge, with the two lines merging at a would-be  multicritical point. The $\chi_P$ heatmap displays a ridge running almost vertically near $\beta\approx1$, confirming that the gauge transition is controlled predominantly by $\beta$. Conversely, the $\chi_H$ ridge bends sharply as $\beta_H$ decreases, terminating in a compact high-susceptibility region at moderate $\beta$ ($\beta_H\approx0.45$--$0.65$), reflecting the $Z_2$ transition's dependence on both couplings. Notably, $\chi_P$ is largely insensitive to the Higgs transition, and $\chi_H$ shows no clear resolved feature at the gauge transition. However, the $Z_2$ transition is visible in both heatmaps. Trajectory (I) crosses both ridges in close succession at small $\beta_H$, whereas trajectory (II) crosses the joint $Z_2$ ridge at large $\beta_H$ and later intersects the Higgs ridge.
\begin{figure*}[t]
    \centering
    \includegraphics[width=\textwidth]{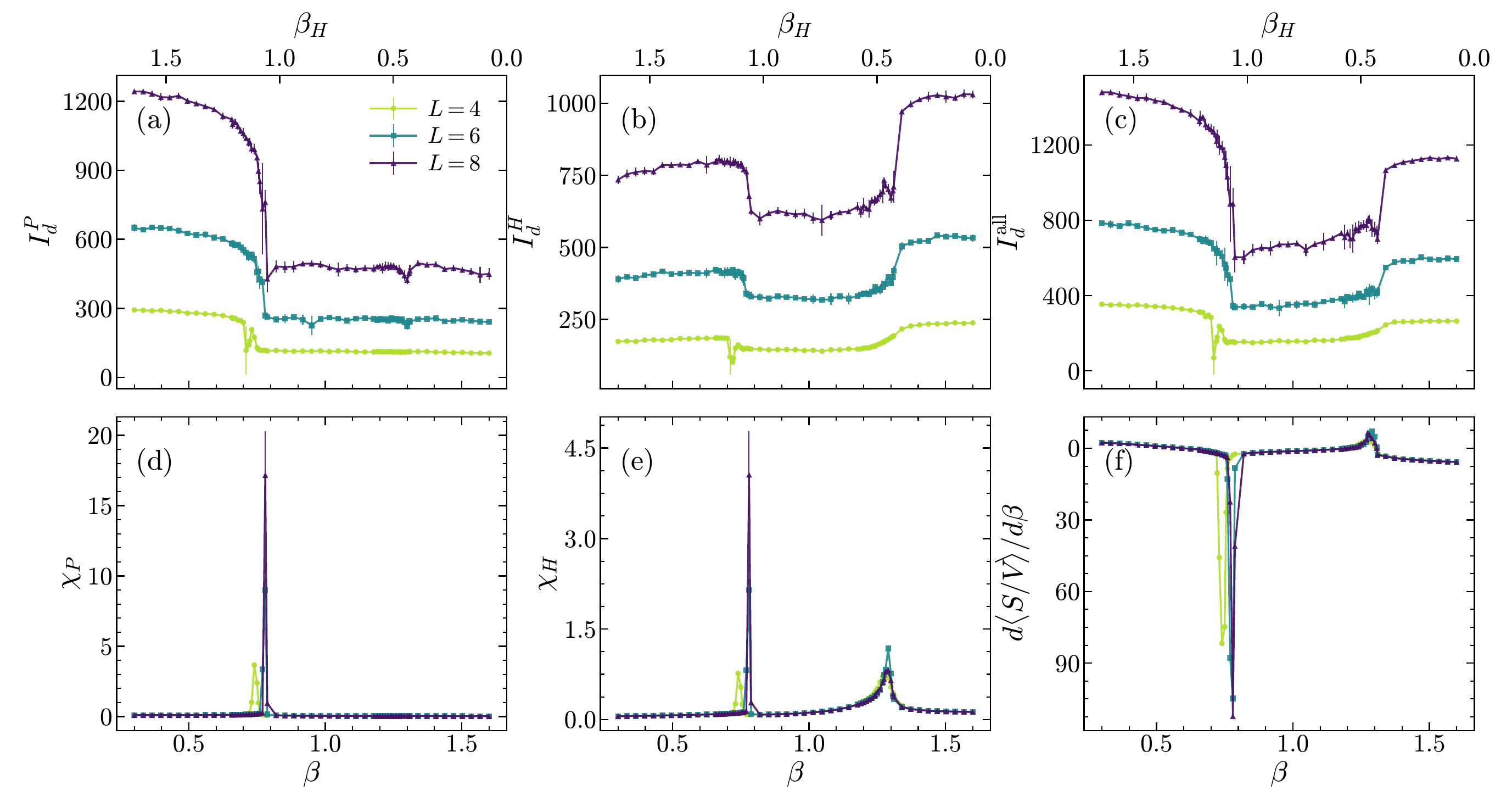}
    \caption{$U(1)$ Higgs model along trajectory (II), $\beta_H=-1.2\,\beta+2$, for $L=4,6,8$. Panels are arranged as in \cref{fig:u1higgs_line1}. (a) to (c), the plaquette, Higgs, and combined intrinsic dimensions. (d) to (f), the plaquette susceptibility, Higgs susceptibility, and the derivative of the mean action density.}
    \label{fig:u1higgs_line2}
\end{figure*}
All scans in this section are performed on $L^4$ lattices ($L=4,6,8$), utilizing $N_{\rm cfg}=5\times10^4$ configurations per coupling point (comprising $10^4$ configurations from each of five independent runs to ensure robust statistical error estimation). Along each trajectory, we evaluate the channel-specific intrinsic dimensions ($I_d^{P}$, $I_d^{H}$, $I_d^{\mathrm{all}}$) and contrast them with conventional thermodynamic observables ($\chi_P$, $\chi_H$, and $d\langle S/V\rangle/d\beta$).

\cref{fig:u1higgs_line1} presents the results for trajectory (I). The plaquette channel intrinsic dimension $I_d^P$ [panel (a)] remains large in the confinement phase. As $\beta$ increases, it shows a sharp drop and a slow partial recovery, forming a pronounced dip. At higher $\beta$, it descends again, settling onto a low plateau. This initial dip lies in the vicinity of  the gauge transition identified by the sharp peak in $\chi_P$ [panel (d)] and by the deep minimum in $d\langle S/V\rangle/d\beta$ [panel (f)], consistently across all system sizes. In contrast, the Higgs channel $I_d^H$ [panel (b)] remains featureless at this point, staying flat across the gauge transition and only descending at the second crossing, precisely where $\chi_H$ [panel (e)] peaks. Unlike the dip observed at the gauge transition, the Higgs transition simply manifests as a sudden step-like drop in $I_d^H$. While the Higgs transition in the four-dimensional charge-two U(1) Higgs model is known to be of first order~\cite{Bowler:1981cj,Callaway:1981rt}, finite-size effects on our current volumes smear it into a continuous crossover (see Appendix~\ref{app:hist}); nevertheless, the sudden geometrical collapse in $I_d$ retains a robust signature of the underlying phase boundary. This constitutes a direct demonstration of channel selectivity: $I_d^H$ is primarily sensitive to the matter transition and exhibits no clearly resolved signature at the gauge transition, whereas $I_d^P$ responds predominatly to the gauge transition. The combined channel $I_d^{\rm all}$ [panel (c)] naturally inherits both structural features. Increasing $L$ from 4 to 8 sharpens all geometric features without altering the selectivity paradigm.

\cref{fig:u1higgs_line2} displays the measurements along trajectory (II), where the two transition lines are well separated. At the first crossing (the $Z_2$ transition), both channels exhibit a strong, synchronized response. The plaquette dimension $I_d^P$ drops abruptly to a low plateau, while the Higgs dimension $I_d^H$ carves out a sharp dip. This joint response is physically expected, as the $Z_2$ gauge transition inherently involves the simultaneous reorganization of both gauge and matter fields, corroborated by coincident peaks in $\chi_P$ and $\chi_H$. However, the behavior at the second crossing (the Higgs transition) provides the most stringent test of channel selectivity. The Higgs channel clearly detects this boundary, with $I_d^H$ rising significantly and $\chi_H$ developing a well-resolved second peak. By contrast, the plaquette channel $I_d^P$ remains dormant on its plateau for $L=4$ and $L=6$, matching the featureless behavior of $\chi_P$. At $L=8$, a faint residual dip emerges in $I_d^P$, suggesting that the gauge sector regains a marginal sensitivity to the matter-driven transition at larger volumes. Thus, a phase transition driven almost exclusively by the matter sector can be essentially invisible to the geometric dimension of the gauge manifold, only being recovered when the relevant scalar degrees of freedom are explicitly included in the metric.

A closer inspection of both trajectories shows small offsets in $\beta$
between the most prominent features in $I_d$ and the peaks of the
conventional susceptibilities. Such offsets are not unexpected, since the two
quantities probe different properties of the finite-volume ensemble. A
susceptibility $\chi$ characterizes fluctuations of a selected bulk
observable, whereas $I_d$ characterizes the local geometry of the joint
distribution of the variables entering the chosen channel. Their corresponding
pseudo-critical locations therefore need not coincide at finite volume,
particularly when the transition region is broadened by finite-size effects.
Nevertheless, the characteristic features of $I_d$ and the susceptibilities
occur in the same transition regions, indicating that they provide
complementary information about the reorganization of the system across the
transition.

Among the transitions probed, the $Z_2$ transition on trajectory (II)
exhibits the clearest first-order characteristics at these system
sizes. As shown in the per-configuration distributions
(\cref{fig:histA1P,fig:histA1H,fig:histA2P,fig:histA2H} in Appendix~\ref{app:hist}),
the gauge and Higgs transitions are smeared into broad crossovers that shift
with $L$, whereas the $Z_2$ transition maintains a
persistent bimodal structure across all studied volumes.
This finite-volume phase-coexistence signal motivates a
finite-size-scaling (FSS) analysis of the $Z_2$ boundary.

\begin{figure}[htbp!]
    \centering
    \includegraphics[width=0.8\linewidth]{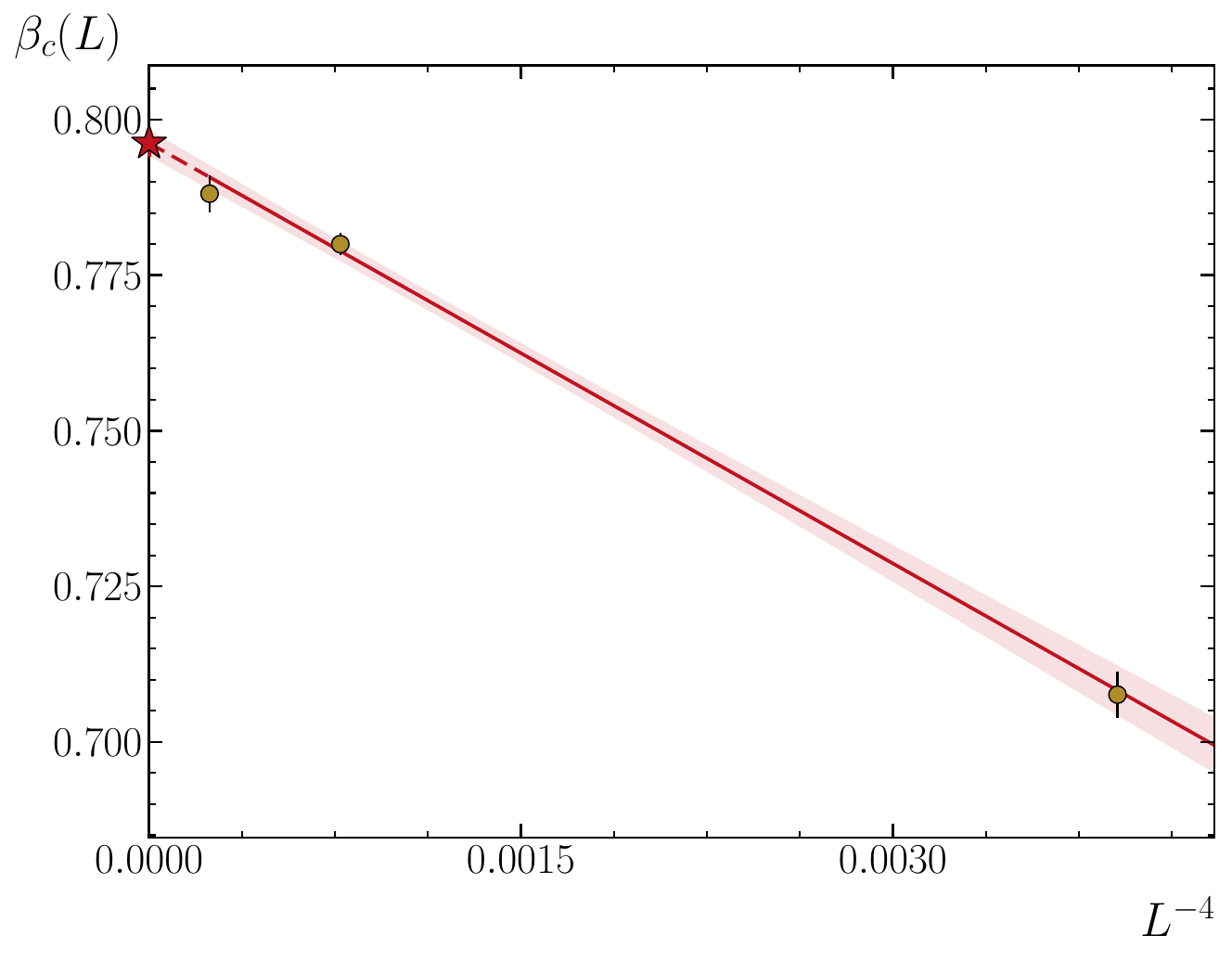}
    \caption{Finite-size scaling of the pseudo-critical coupling $\beta_c(L)$
identified along trajectory (II) at the $Z_2$ transition. For each lattice
size $\beta_c(L)$ is the position of the minimum of $I_d^{\rm all}$, obtained
from the cusp fit described in Appendix~\ref{app:cusp}, and its error bar is a
parametric bootstrap estimate. Having identified this transition as first order
from the persistent bimodal distributions of \cref{app:hist}, we use the
finite-size shift expected in that case, $\beta_c(L)=\beta_c(\infty)+a\,L^{-4}$,
to extrapolate to infinite volume. The solid line is a weighted linear fit to
the three data points and the shaded region is its $1\sigma$ band, with the
dashed segment indicating where the line is extrapolated beyond the largest
lattice size. The star at $L^{-4}=0$ is the intercept, whose uncertainty is
propagated from the fit covariance, and it gives
$\beta_c(\infty)=0.7963(22)$.}
    \label{fig:fss}
\end{figure}

\cref{fig:fss} shows the finite-size scaling of the pseudo-critical coupling
$\beta_c(L)$ associated with the $Z_2$ transition. For each lattice size,
$\beta_c(L)$ is identified with the position of the minimum of
$I_d^{\rm all}$ along trajectory (II), which we determine by fitting the scan
in a window bracketing the minimum with a smooth polynomial background
supplemented by a two-sided power-law term whose exponent is left free. 
The form is motivated by the standard power-law description of
cusp singularities near critical points \cite{Cardy:1996}, and is
used here only as a phenomenological interpolation for locating
the sharp minimum, and it is better suited to the sharp dip found
at the $Z_2$ transition than a polynomial interpolation alone, because the
position of the minimum enters as a fit parameter rather than as a stationary
point reconstructed from the balance of polynomial coefficients across the
whole window. The uncertainty of $\beta_c(L)$ is obtained from a parametric
bootstrap over the error bars of $I_d^{\rm all}$, and the procedure is
described in Appendix~\ref{app:cusp}. The resulting values are consistent with
an approximately linear dependence on the inverse four-volume $L^{-4}$, as
expected for the leading finite-size shift of a first-order transition in
$d=4$ dimensions, and the extrapolation yields
$\beta_c(\infty)=0.7963(22)$.
Given the limited range of lattice sizes, this scaling should be
regarded as supporting the first-order interpretation suggested by the
bimodal distributions rather than establishing it independently.
A more quantitative finite-size analysis of the gauge and Higgs transitions
will require simulations on substantially larger lattices and is left for
future work.

\section{Discussion}
\label{sec: discussion}

The distinct geometric behaviors of the intrinsic dimension ($I_d$) observed across our benchmark systems warrant a careful physical interpretation, particularly when positioned as a methodological foundation for probing the finite-temperature crossover structure of full QCD.

In conventional spin systems, such as the two-dimensional clock models investigated in \cref{sec: results_clock}, the phase transition is fundamentally driven by the competition between thermal fluctuations and magnetic ordering. At low temperatures, the system settles into an ordered state where field configurations are tightly correlated, restricting the local geometry explored by the sampled ensemble. As the temperature increases, the proliferation of thermal noise forces the system to sample a highly disordered, isotropic configuration space. Geometrically, this reorganization of correlations is reflected in a pronounced overall increase of $I_d$ from the ordered to the disordered regime. This trend is not strictly monotonic, however: as detailed below, the emergent quasi-critical phase interrupts it by sustaining a stable low-$I_d$ geometric regime over a finite temperature window.

A mathematically analogous, yet physically distinct, logic governs the four-dimensional $U(1)$ Higgs model when tuning the bare gauge coupling $\beta = 1/g^2$. In this isotropic four-dimensional Euclidean lattice system, the low-$\beta$ (strong coupling) regime is dominated by non-linear gauge fluctuations and the proliferation of topological defects. This strongly fluctuating configuration ensemble yields a large $I_d^P$ in the confinement phase. Conversely, transitioning to the Coulomb phase at high $\beta$ suppresses these non-linearities, reducing the non-Gaussian gauge-field fluctuations and reorganizing the local geometry of the sampled ensemble, which is reflected in a lower $I_d^P$.

A conceptual tension emerges when extending this geometric diagnostic to the finite-temperature phase diagram of full QCD in 3+1 dimensions. In full QCD, the system is governed by the physical temperature $T$, and the effective interaction strength is dynamically modulated by the running coupling constant $g(T)$ via asymptotic freedom. Unlike the spin models (where heating increases $I_d$) or the four-dimensional $U(1)$ Higgs model studied here (where weak coupling decreases $I_d$), full QCD near the crossover region ($T \sim T_c$) may involve competing effects on the local geometry of the sampled configuration ensemble:
\begin{itemize}
    \item \textbf{Thermal Liberation of Degrees of Freedom.} As $T$ surpasses the pseudo-critical crossover temperature, confinement flux tubes melt and quark and gluon degrees of freedom become thermally active in the quark-gluon plasma (QGP). The associated reorganization of correlations and effective constraints may increase the number of locally independent directions explored by the ensemble, and hence tend to increase $I_d$.
    \item \textbf{Asymptotic Freedom.} Concurrently, higher physical temperatures translate to shorter length scales, where the strong coupling weakens. At asymptotically high temperatures, the system approaches the Stefan-Boltzmann limit of a weakly interacting gas. The resulting suppression of strong non-linear and non-Gaussian correlations may reduce the effective local complexity of suitably chosen gauge-invariant configuration-space channels, and hence tend to decrease $I_d$.
\end{itemize}
Since the 2NN estimator depends on the ratio $r_2/r_1$, a uniform rescaling of configuration-space distances does not by itself change $I_d$. Any temperature dependence must therefore arise from a reorganization of the local probability distribution, such as changes in correlations, effective constraints, anisotropies, or non-Gaussian structure. The sign and detailed behavior of $I_d$ in full QCD consequently cannot be inferred a priori, but may reflect the interplay of these competing effects.

Beyond serving as a numerical testing ground for closely spaced transitions, the $q$-state clock model offers a useful methodological comparison with QCD, particularly regarding extended strongly correlated regimes. For $q > 4$, the clock model accommodates a quasi-long-range-ordered(QLRO) phase bounded by two distinct Berezinskii-Kosterlitz-Thouless (BKT) transitions. In this intermediate phase, the system exhibits critical scale invariance across a continuous finite range of temperatures rather than at an isolated singular point.

Although the clock model and finite-temperature QCD differ fundamentally in their microscopic dynamics and universality properties, a qualitative analogy may be drawn with proposed intermediate regimes in finite-temperature QCD and many-flavor gauge theories: the existence of a semi-QGP regime, characterized by a residual center-symmetry structure that lifts explicit linear confinement while preserving non-perturbative correlations, or, in a complementary picture, a strongly coupled QGP (sQGP), in which residual non-perturbative color interactions remain sufficiently potent to form liquid-like, highly correlated structures rather than a fully decoupled parton gas. The characteristic geometric valley of $I_d$ observed in the clock model (\cref{sec: results_q5}) demonstrates, within this benchmark system, that the 2NN estimator can identify an extended correlated intermediate regime rather than only isolated transition points. This motivates testing whether analogous geometric information can be extracted in more complex gauge theories.

Furthermore, the discretization of continuous $U(1)$ symmetry into a discrete $Z_q$ subgroup in the clock model shares a discrete-symmetry structure with center symmetry ($Z_N$) in pure $SU(N)$ gauge theories, although the microscopic degrees of freedom and symmetry realizations are different. In full QCD with dynamical fermions, the $Z_3$ center symmetry is explicitly broken by dynamical quarks of finite mass, and the pure-gauge deconfinement transition is replaced by crossover behavior in the physical theory. The $q$-state clock model therefore provides a simple benchmark for examining how nearby transition scales and correlated intermediate regimes are reflected in configuration-space geometry. The resilience of the stable $I_d$ valley in the $q=5$ clock model---where the intermediate phase is severely compressed---suggests that intrinsic dimensions may remain informative when distinct physical scales are closely spaced or when no exact order parameter is available. Establishing whether this diagnostic can disentangle deconfinement and chiral crossover scales in full QCD will require direct calculations with appropriately constructed gluonic and fermionic channels.

\section{Conclusion}
\label{sec: conclusion}

We have shown that the intrinsic dimension ($I_d$) of Monte Carlo
configuration ensembles, estimated via the Two-Nearest-Neighbors
(2NN) method, can serve as an unsupervised geometric diagnostic for
lattice systems with multiple or strongly overlapping phase transitions.
By testing this methodology on two benchmark systems, we highlighted
complementary aspects of its diagnostic capability. In the two-dimensional
$q$-state clock model, $I_d(T)$ successfully distinguishes the ordered,
quasi-critical, and disordered regimes. Crucially, it isolates the intermediate
quasi-long-range-ordered phase as a stable geometric valley even in the
closely spaced $q=5$ case, where conventional energy and magnetization
fail to resolve the two adjacent BKT transitions. In the four-dimensional
$U(1)$ Higgs model, introducing a channel-decomposed $I_d$ revealed
a pronounced sector dependence. The geometric dimension of a given
channel responds most strongly to transitions involving the degrees
of freedom represented in that channel. Consequently, a matter-driven
transition produces little or no resolved signal in the plaquette
channel at the present lattice sizes, while it is clearly detected when the
scalar degrees of freedom are included.

The two benchmarks together suggest a general working procedure. At each value
of the external condition one generates an ensemble of configurations by Monte
Carlo simulation, keeping the number of configurations fixed across the scan and
retaining them at separations well beyond the integrated autocorrelation time.
These ensembles are passed to the 2NN estimator, which returns $I_d$ as a
function of the external condition once a distance between configurations has
been specified. The shape of the resulting curve then carries the physical
information, since the minima and the abrupt changes of $I_d$ mark where the
sampled configuration space reorganises and thereby locate the transitions.
Repeating the scan for several lattice sizes finally allows a finite size
scaling analysis of the pseudo-critical values extracted in this way, with the
ansatz chosen according to the nature of the transition. Note that for gauge system, the estimator must be applied to
gauge invariant variables constructed from the field configurations, since
otherwise the nearest neighbour structure is dominated by displacements along
the gauge orbit and what is measured is the intrinsic dimension of that orbit
rather than the geometry of the physical dynamics, as demonstrated in
\cref{app:gauge_redundancy}.

When several transitions are expected within a narrow range, this procedure is
supplemented by the channel decomposition of \cref{sec: method_u1higgs}. The gauge
invariant variables are partitioned into sectors corresponding to physically
distinct degrees of freedom, and the estimator is applied to each sector and to
their union along the same trajectory, so that all curves originate from one and
the same ensemble. A feature in one channel while the other stays flat
identifies a transition driven predominantly by the degrees of freedom of that
channel, whereas a simultaneous response signals a transition across which both
sectors reorganise together. The union plays the role of a no preselection limit
and is the natural starting point when the relevant degrees of freedom are not
known in advance. Two channels responding at nearby but distinct values of the
external condition then indicate two separate transition scales even where a
single bulk observable shows only one broad crossover.

Because $I_d$ is a geometric quantity extracted from the local
structure of the sampled ensemble, it requires no preselected
order parameter, making it potentially useful in scenarios where
exact order parameters are unavailable, such as smooth crossovers
or explicitly broken symmetries. Its interpretation nevertheless
depends on the chosen representation, distance metric, and sampling
properties. This motivates several natural extensions of the framework. Note that since the intrinsic dimension is constructed from the geometry of the sampled data rather than from the symmetries of the underlying theory, and there is no strict correspondence between the two, although a feature in $I_d$ signals a critical transition point, it does not tell us which symmetry is involved or exactly which type of phase transition the system is going through.

The most immediate direction is the extension to non-Abelian gauge theories,
first pure $SU(2)$ and $SU(3)$ and ultimately full QCD with dynamical fermions,
where the channel decomposition can be applied to
gauge-invariant or explicitly gauge-reduced gluonic variables and to
suitably constructed fermionic observables separately
to investigate whether the deconfinement and chiral crossover scales
can be disentangled at physical quark masses~
\cite{Aoki:2006we,Borsanyi:2010bp,Bazavov:2011nk,HotQCD:2018pds,
Borsanyi:2020fev,Ding:2015ona,Ratti:2018ksb}.
A second direction is a systematic comparison with alternative geometric and
topological probes, including other intrinsic-dimension estimators~
\cite{Levina:2004mle,Granata:2016acc} and topological data analysis~
\cite{Carlsson:2009zke,Spitz:2022tul,Sale:2022qfn}, to delineate which features
of the phase structure are robust and which are estimator-dependent.
A third direction concerns regimes where conventional order parameters are
intrinsically elusive, including finite-density lattice gauge theories affected
by the sign problem and out-of-equilibrium dynamics~
\cite{Broecker:2017hjl,Boyda:2022nmh,Aarts:2026uiu};
in such settings, geometric diagnostics may provide complementary
information when suitable ensembles can be sampled, although they do not by
themselves resolve the sign problem. Finally, the same framework can be
extended beyond QCD to disordered and frustrated systems, and to
topological and quantum critical points where no local order parameter is
available~\cite{Mendes-Santos:2020msa,Mendes-Santos:2021mhs}.

\section*{Note}
\paragraph{Code availability}
An open-source code with scripts reproducing all results of this work will be made publicly available in the next public version of this work.
\paragraph{Use of AI tools}
Claude Opus 4.8/Fable 5 (Anthropic) and Gemini Pro 3.1 (Google) assisted substantially with numerical cross-checks and with polishing the manuscript. All derivations, results, and statements were verified by the authors, who take full responsibility for the content.

\begin{acknowledgments}
We thank Gert Aarts, Chris Allton, Ryan Bignell, Heng-Tong Ding, Kenji Fukushima, Jeffrey Giansiracusa, Biagio Lucini, Jan Pawlowski, Alexander Rothkopf, Anthony Francis for insightful discussions.
We thank the DEEP-IN working group at RIKEN-iTHEMS for support in the preparation of this paper.
JM is supported by the China Postdoctoral Science Foundation under Grant No. 2026M793760. 
TH is supported by the JSPS KAKENHI Kiban-S Grant No. 26K21721 and the Japan Science and Technology Agency (JST) as part of Adopting Sustainable Partnerships for Innovative Research Ecosystem (ASPIRE),  Grant No. JPMJAP2318.
MH is supported in part by the National Natural Science Foundation of China (NSFC) Grant Nos: 12235016, 12221005. LW is supported by the JSPS KAKENHI Grant No. 25H01560, and JST-BOOST Grant No. JPMJBY24H9. LW and TH are also supported by the RIKEN-TRIP initiative (RIKEN-Quantum).

\end{acknowledgments}

\appendix

\section{Linear scaling of the empirical CDF}
\label{app:linear}

\begin{figure}[htbp!]
    \centering
    \includegraphics[width=0.5\textwidth]{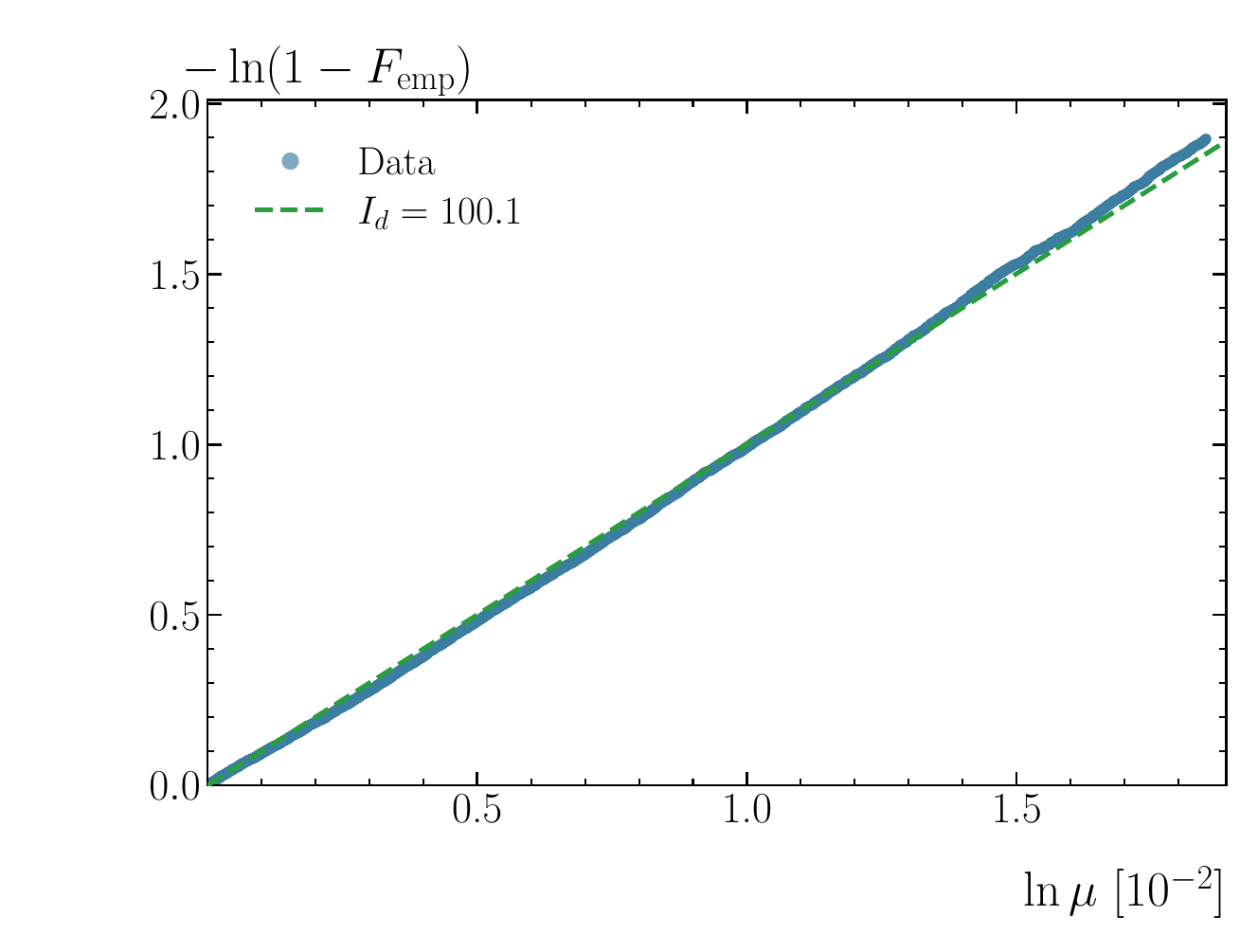}
    \caption{
    Results of the 2NN method. Linear scaling of the empirical CDF conforming to \cref{eq:linear_relation} for the $q=9$ clock model at $T=0.5$, using $L=60$ and $10^{4}$ configurations. The fitted slope yields an intrinsic dimension $I_d\approx 100$.}
    \label{fig:2nn_example}
\end{figure}

\begin{figure*}[htbp!]
\centering
\includegraphics[width=\textwidth]{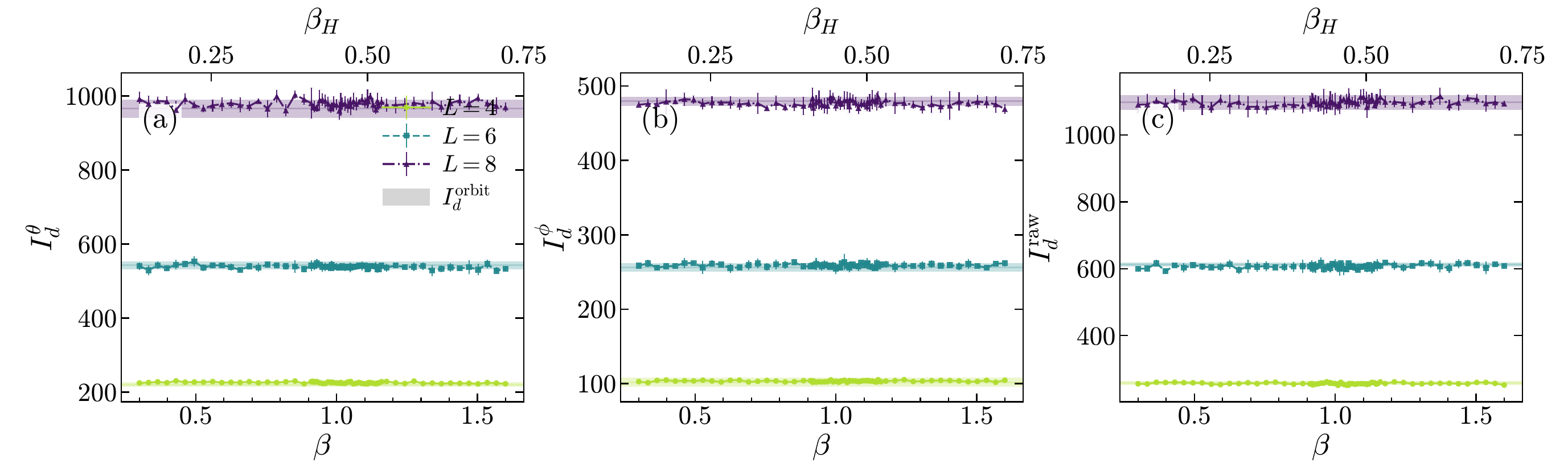}
\caption{Raw intrinsic dimension $I_d^{\text{raw}}$ evaluated directly on the
gauge-variant field variables along trajectory (I), $\beta_H = 0.45\beta$, for
$L = 4, 6, 8$: (a) the gauge-link phases $\{\theta_{n,\mu}\}$, (b) the Higgs
phases $\{\varphi_n\}$, and (c) their union. At each point
$N_{\rm cfg} = 4000 \times 5$ configurations are used. The pale horizontal band in
each panel is the intrinsic dimension of a pure gauge orbit,
$I_d^{\text{orbit}}(L)$, measured on $20{,}000$ random gauge copies of a
single configuration and drawn in the colour of the corresponding $L$ curve;
the band half width is its statistical error.}
\label{fig:raw_ID_traj1}
\end{figure*}

Here we demonstrate the robustness of the 2NN estimator. \cref{fig:2nn_example} illustrates the empirical cumulative distribution conforming to the scaling relation in \cref{eq:linear_relation}, generated at $T=0.5$ for the $q=9$ model ($L=60$, $10^{4}$ configurations). The excellent linear collapse strictly through the origin validates the local uniform density assumption, yielding $I_d \approx 100$. Comparable fitting quality is consistently maintained across all temperatures and system sizes studied.

\section{Gauge Redundancy and the Role of Gauge-Invariant Channels}
\label{app:gauge_redundancy}

In \cref{sec: method_u1higgs}, the intrinsic dimension ($I_d$) of the
four-dimensional $U(1)$ Higgs model was evaluated strictly on gauge-invariant
channels. A fundamental question is whether $I_d$ evaluated directly on the
raw, gauge-variant field variables $(\{\theta_{n,\mu}\}, \{\varphi_n\})$ can
detect phase transitions. Here we show that in the present setup the
raw-field $I_d$ shows no resolved transition signal and is instead dominated
by gauge redundancy.

As shown in \cref{fig:raw_ID_traj1}, $I_d^{\text{raw}}$ along trajectory
(1) shows no clearly resolved feature at either the gauge or the
Higgs transition across $L = 4, 6, 8$, in each of the three raw channels.
Instead of exhibiting transition signatures, it remains close to a
size-dependent plateau with extremely small error bars.

To identify the origin of this behavior, we isolate the geometry of pure gauge
transformations. In lattice gauge theory, gauge-equivalent
configurations form gauge orbits under
$\mathcal{G}=\prod_n U(1)_n$, while physical configurations are represented
by equivalence classes in the quotient $\mathcal{C}/\mathcal{G}$. We take a
single equilibrium configuration $C_0$ and apply independent random local
transformations $\Omega_n=e^{i\alpha_n}$, with $\alpha_n$ uniform on
$[0,2\pi)$, to generate $N_{\text{gauge}}=20{,}000$ configurations that are
strictly gauge-equivalent to $C_0$ and therefore
represent the same physical state. This dataset is passed through
exactly the same pipeline as the physical ensembles, and the resulting
intrinsic dimension of pure gauge redundancy,
$I_d^{\text{orbit}}(L)$, is shown as the pale bands in
\cref{fig:raw_ID_traj1}. Every raw curve lies inside its own band over the
whole trajectory, including at both transitions. On the same dataset the
gauge-invariant channels are exactly degenerate, since all $20{,}000$
configurations share identical $P_{\mu\nu}(n)$ and $H_\mu(n)$: the invariant
channels see the gauge orbit as a single point, as they should.

The interpretation is then natural. The local gauge group contains
$V=L^4$ continuous gauge parameters. For the full gauge--Higgs
configuration, a generic gauge orbit has continuous dimension $V$ up to
discrete stabilizers, whereas for the gauge-link variables alone one global
constant transformation acts trivially, leaving $V-1$ independent continuous
gauge directions. The physical fluctuations that drive the transitions live
on the quotient $\mathcal{C}/\mathcal{G}$. Because the 2NN estimator is built
from the local neighbor distances $r_1$ and $r_2$, and because
the action and hence the Boltzmann weight are constant along each
gauge orbit, the observed agreement with the pure-orbit data
indicates that the nearest-neighbor structure in raw field space is dominated
by displacements along the orbit. What $I_d^{\text{raw}}$
primarily reflects is therefore the local geometry of gauge-orbit
directions rather than the correlation structure of the physical dynamics.

That $I_d^{\text{orbit}}(L)$ itself falls below
the corresponding formal orbit dimension at $L=6,8$ reflects
finite-sample bias or saturation of the 2NN estimator at large
intrinsic dimension, rather than a property of the gauge group.
Because the raw and pure-orbit datasets are analyzed with the same
estimator and metric, their close agreement supports the interpretation
above, although the absolute value of $I_d$ at finite sample size should not
be identified with the exact orbit dimension.
Removing gauge redundancy is therefore necessary before an
intrinsic-dimension analysis of field-space configurations can be given a
direct physical interpretation; gauge-invariant channels provide a natural
way to achieve this.

\begin{figure*}[htbp!]
\centering
\includegraphics[width=\textwidth]{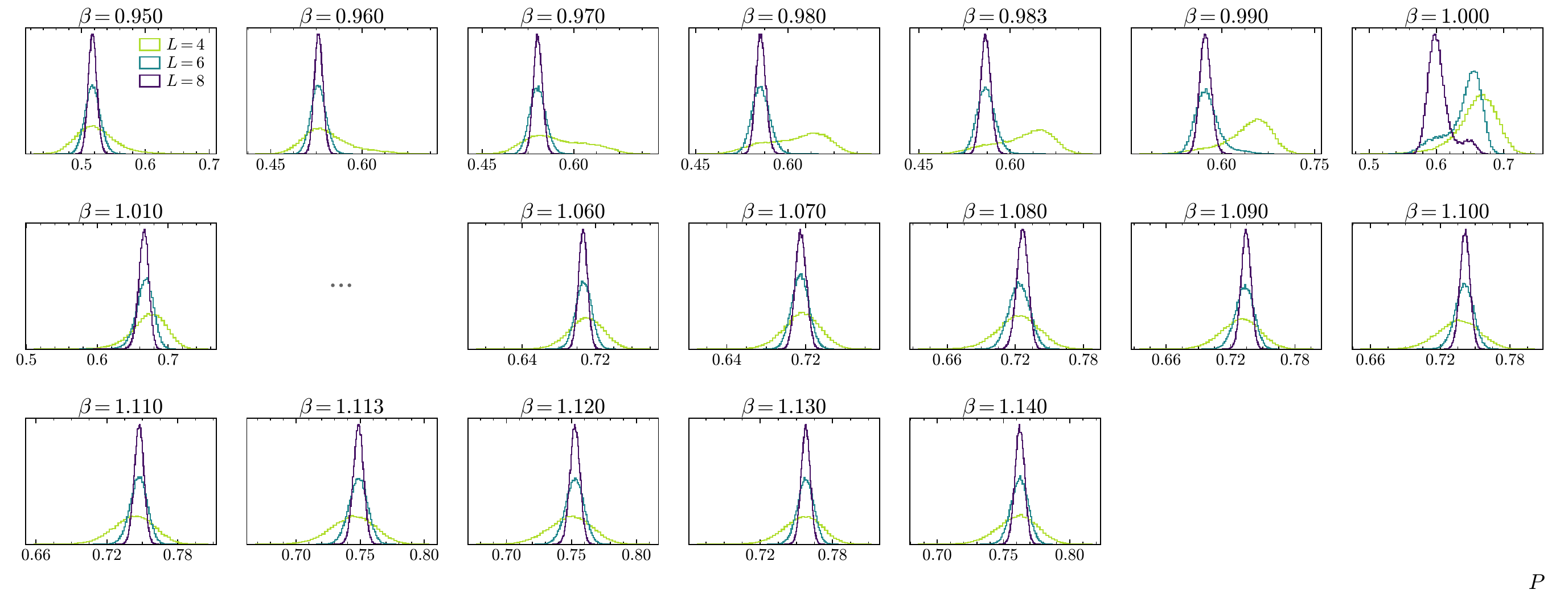}
\caption{Per-configuration distribution of the plaquette channel
variable $P$ versus $\beta$ along trajectory (I), for $L = 4, 6, 8$,
shown only in the two windows bracketing the gauge transition
($\beta = 0.950$--$1.010$) and the Higgs transition
($\beta = 1.060$--$1.140$). The omitted scan range in between is
unimodal throughout and is collapsed into a single `$\cdots$''
panel.}
\label{fig:histA1P}
\end{figure*}

\begin{figure*}[htbp!]
\centering
\includegraphics[width=\textwidth]{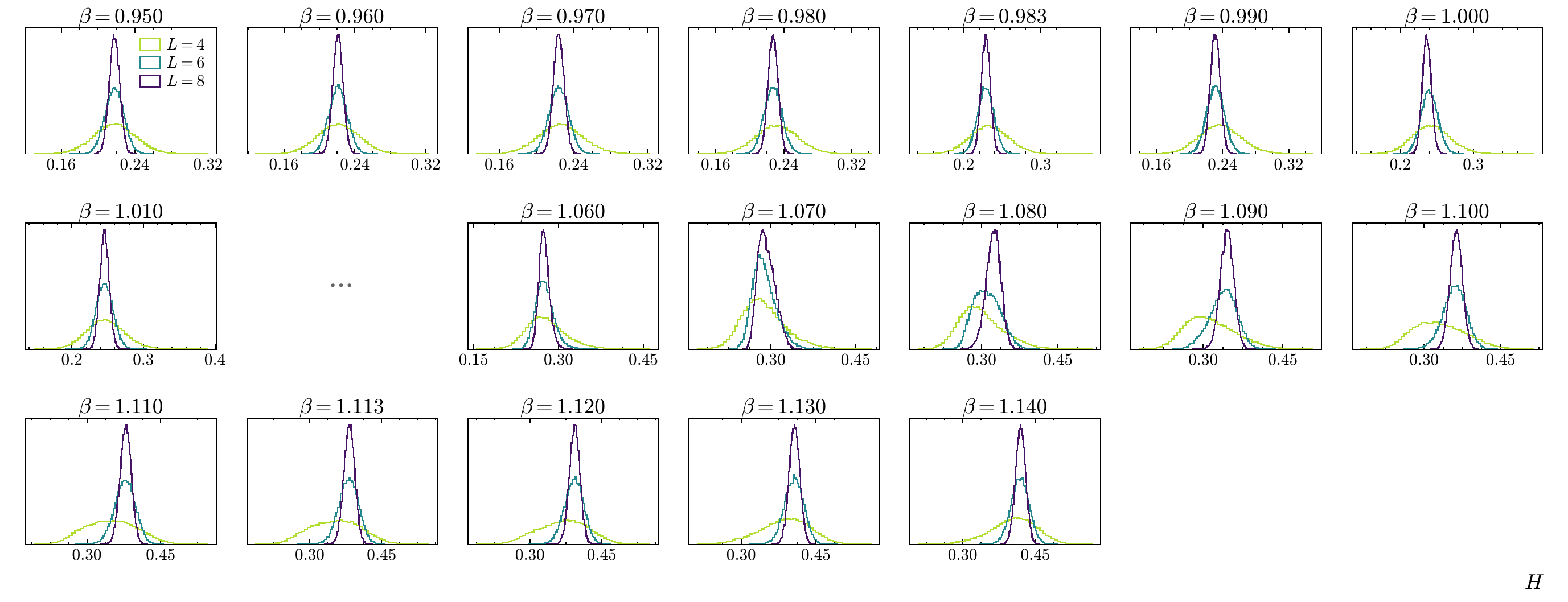}
\caption{Per-configuration distribution of the Higgs channel
variable $H$ versus $\beta$ along trajectory (I), for
$L = 4, 6, 8$, shown in the same two windows as
\cref{fig:histA1P}. The omitted range between the windows is
unimodal and is marked by `$\cdots$''.}
\label{fig:histA1H}
\end{figure*}

\begin{figure*}[htbp!]
\centering
\includegraphics[width=\textwidth]{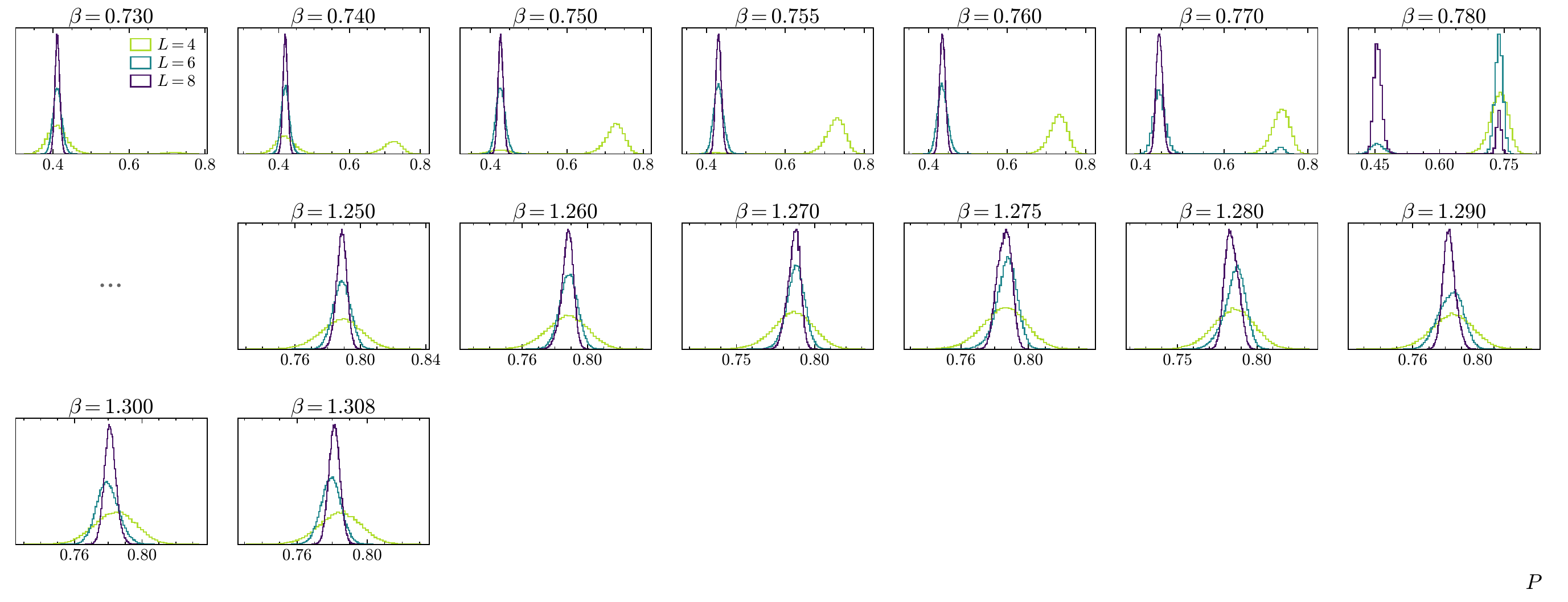}
\caption{Per-configuration distribution of the plaquette channel
variable $P$ versus $\beta$ along trajectory (II), for
$L = 4, 6, 8$, shown only in the two windows bracketing the $Z_2$ transition ($\beta = 0.730$--$0.780$) and the Higgs transition ($\beta = 1.250$--$1.308$). The omitted range in between is marked by `$\cdots$''.}
\label{fig:histA2P}
\end{figure*}

\begin{figure*}[htbp!]
\centering
\includegraphics[width=\textwidth]{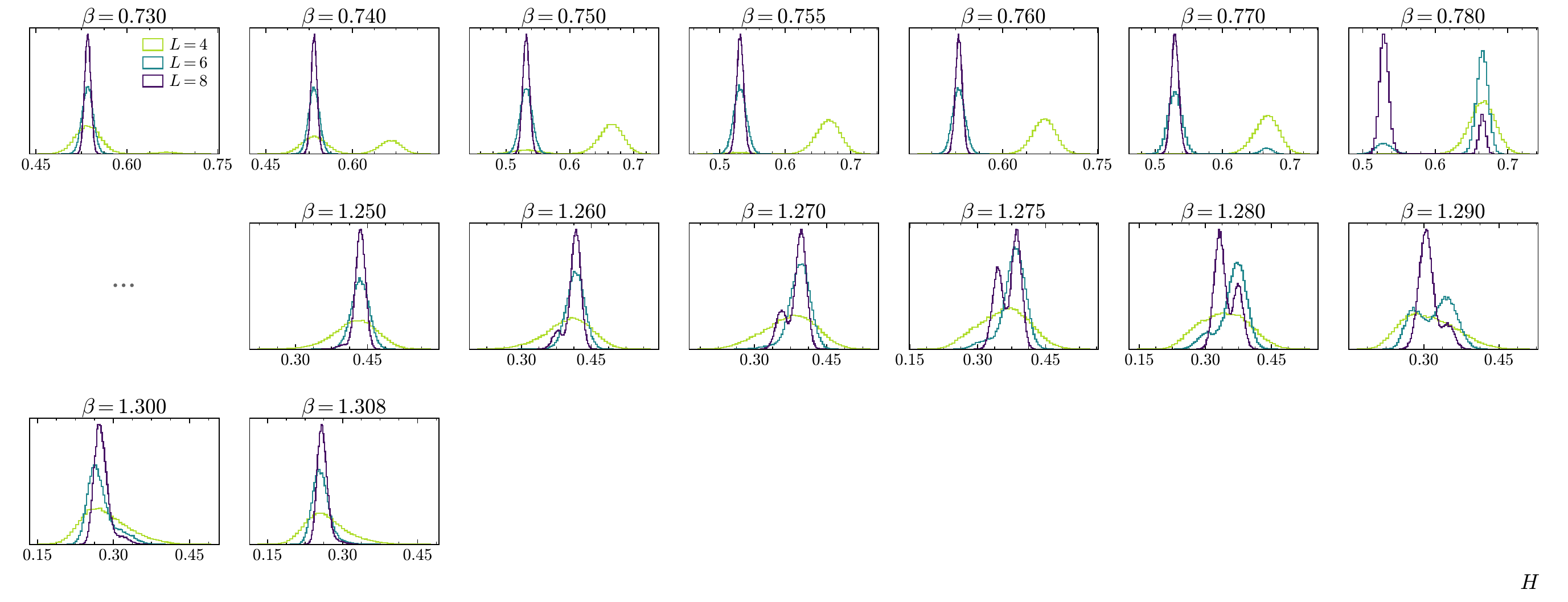}
\caption{Per-configuration distribution of the Higgs channel
variable $H$ versus $\beta$ along trajectory (II), for
$L = 4, 6, 8$, shown in the same two windows as
\cref{fig:histA2P}. The omitted range in between is marked
by `$\cdots$''.}
\label{fig:histA2H}
\end{figure*}

\section{Per-Configuration Distributions near the Transitions
of the $U(1)$ Higgs Model}
\label{app:hist}

To assess where finite-size effects compromise a controlled
finite-size-scaling analysis, we show here the per-configuration
distributions of the plaquette channel variable $P$ and the Higgs
channel variable $H$, in the neighborhood of the transitions crossed
by trajectory (I) and trajectory (II), for $L=4,6,8$. Away from these
neighborhoods every scan point gives an ordinary single-peaked
distribution, so only the transition regions are shown in the figures
below and the remaining points are omitted. At some scan points the
distribution for at least one lattice size develops two coexisting
peaks rather than a single one, which is
a finite-volume signature of phase coexistence and is
suggestive of first-order behavior. Whether these two peaks
stay well separated as $L$ grows or instead merge back into a broad
single peak is what determines whether a controlled finite-size
scaling analysis can be carried out for a given transition.

\subsection{Trajectory (I), $\beta_H = 0.45\beta$}

\cref{fig:histA1P} and \cref{fig:histA1H} show the
per-configuration distributions of the plaquette and Higgs channel
variables along trajectory (I), for $L=4,6,8$. For the plaquette
channel, the distribution near the first transition encountered, the
gauge transition, broadens for $L=4$ and its peak drifts as $\beta$
increases through the transition region. As $L$ grows this broadening
develops into a shoulder on the main peak, and by $L=8$ the shoulder
has grown into a distinct secondary peak,
suggesting emerging finite-volume phase coexistence.
Near the second transition, the Higgs transition, the plaquette
distribution remains single peaked at every $L$ shown, so this
channel shows no clear sensitivity to it at these lattice
sizes. The Higgs channel shows the complementary pattern. Its
distribution shows no visible change at the gauge
transition and stays single peaked there throughout, while near the
Higgs transition it broadens noticeably without ever developing a
clearly resolved double peak at any of the lattice sizes studied,
consistent with the plateau-like drop of $I_d^H$ at the Higgs
transition discussed in \cref{sec: Results} rather than the
sharper dip that would accompany a cleanly resolved coexistence.
Because of this the order of the Higgs transition cannot be settled
from the $H$ channel alone at the present lattice sizes, and doing so
would require simulations on larger lattices.

\subsection{Trajectory (II), $\beta_H = -1.2\beta + 2$}

\cref{fig:histA2P} and \cref{fig:histA2H} show the same
distributions along trajectory (II), for $L=4,6,8$. For the plaquette
channel, the distribution near $\beta\approx0.730$ to $0.780$,
corresponding to the $Z_2$ transition, develops a clear double peak
at all three lattice sizes, with the two peaks well separated rather
than merging as $L$ increases.
This persistent bimodal structure is consistent with a
first-order transition and motivates the finite-size-scaling analysis
used to extrapolate $\beta_c(\infty)$ in \cref{fig:fss}.
Near the Higgs transition, at
$\beta\approx1.250$ to $1.308$, the plaquette distribution shows no clear tendency toward a double peak at any lattice size,
though the small, size-dependent feature seen in $I_d^P$ near this
transition and discussed in \cref{sec: Results} suggests the plaquette
channel still carries a weak residual response to it that the raw
distribution alone does not resolve. The Higgs channel shows the same
well-separated double peak at the $Z_2$ transition as the plaquette
channel, in agreement with the picture already established there.
Near the Higgs transition, however, the $H$ distribution does develop
two peaks, but they remain incompletely separated at every lattice
size studied, suggesting finite-volume phase coexistence but
not allowing the thermodynamic order of the transition to be
established from the present lattice sizes alone. Extending the
finite-size-scaling analysis to the transitions other than the $Z_2$
transition along trajectory (II) will require simulations on larger
lattices.

\section{Location of the Minima of $I_d$ and the Fit to the Cusp Form}
\label{app:cusp}

This appendix describes how the pseudo-critical couplings $\beta_c(L)$ used in
the finite-size-scaling analysis of the $Z_2$ transition of the $U(1)$ Higgs
model, \cref{fig:fss}, are extracted from the scans of $I_d^{\rm all}$ along
trajectory (II). It does not apply to the clock model of
\cref{sec: results_clock}, where the minima of $I_d(T)$ are broad and shallow
and the form given below is not constrained by the data. The pseudo-critical
temperatures quoted there are obtained instead from a weighted quartic fit over
a temperature window bracketing the minimum, with the uncertainty of $T^{*}$
estimated by the same parametric bootstrap described at the end of this
appendix.

At the $Z_2$ transition the combined channel intrinsic dimension drops over a
range of $\beta$ comparable with the spacing of the scan, so a smooth
interpolation locates the minimum only through the balance of its coefficients
over the whole fit window. We therefore fit $I_d^{\rm all}$ in a window
bracketing the dip with a polynomial background supplemented by a two-sided
power law,
\begin{equation}
I_d(\beta)=\sum_{k=0}^{2}a_k\,(\beta-\beta_{\rm th})^{k}+
\begin{cases}
C_-\,(\beta_{\rm th}-\beta)^{p}, & \beta<\beta_{\rm th},\\[2pt]
C_+\,(\beta-\beta_{\rm th})^{p}, & \beta>\beta_{\rm th},
\end{cases}
\label{eq:cusp}
\end{equation}
which is the form is motivated by the standard power-law description of cusp singularities near critical points \cite{Cardy:1996}. For $0<p<1$ and $C_\pm>0$ the derivative of \cref{eq:cusp} diverges with opposite signs on the two sides of $\beta_{\rm th}$, so that $\beta_{\rm th}$ is the position of the minimum itself and is obtained directly as a parameter of the fit.

At fixed $\beta_{\rm th}$ and $p$ the remaining parameters enter linearly and
are obtained by weighted least squares with weights given by the inverse
squared errors of $I_d$. The two nonlinear parameters are determined by
scanning $\beta_{\rm th}$ across the fit window and $p$ over the interval
$0<p<1$, so that the result does not depend on a starting guess. Values of $p$
close to unity are excluded, since the power-law term then becomes degenerate
with the linear background.

The uncertainty of $\beta_{\rm th}$ is estimated by a parametric bootstrap in
which the values of $I_d$ are resampled according to their error bars and the
complete fit is repeated for each sample. We take the standard deviation of the
resulting distribution as the error bar of $\beta_c(L)$, and we have checked
that the interval obtained from the profiled $\chi^2$ of the fit is compatible
with it. The three values of $\beta_c(L)$ are then extrapolated with a weighted
linear fit in $L^{-4}$, with the covariance rescaled by $\chi^2$ per degree of
freedom and propagated to the intercept, which yields
$\beta_c(\infty)=0.7963(22)$ with $\chi^{2}$ per degree of freedom of $1.2$.

\bibliography{reference}
\end{document}